\documentclass[twocolumn,nofootinbib,showpacs,prl,aps]{revtex4-2}
\usepackage{epsf, epsfig,graphicx}
\usepackage{subfigure}
\usepackage{bm}
\usepackage{amsmath}
\usepackage{amssymb}
\usepackage{color}

\begin{document}
\title{Mechanosensitive bonds induced complex cell motility patterns}
\author{
  Jen-Yu Lo$^{1}$, Yuan-Heng Tseng$^{1}$ and Hsuan-Yi Chen $^{1,2,3}\footnote{email address: hschen@phy.ncu.edu.tw}$
}
\affiliation{
  $^{1}$Department of Physics, National Central University, Jhongli 32001, Taiwan\\
  $^{2}$Institute of Physics, Academia Sinica, Taipei, 11529, Taiwan \\
  $^{3}$Physics Division, National Central for Theoretical Sciences, Taipei, 10617, Taiwan
}	 	
\date{\today}
\begin{abstract}
The one-dimensional crawling movement of a cell is considered in this theoretical study. 
Our active gel model shows that 
for a cell with weakly mechanosensitive adhesion complexes, as myosin contractility increases, a cell starts to move at a constant velocity. 
As the mechanosensitivity of the adhesion complexes increases, 
a cell can exhibit stick-slip motion.  
Finally, a cell with highly mechanosensitive adhesion complexes exhibits periodic back-and-forth migration. 
A simplified model which assumes that the cell crawling dynamics are controlled by the evolution of the myosin density dipole and the asymmetry of adhesion complex distribution 
captures the motility behaviors of crawling cells qualitatively.  
It suggests that the complex cell crawling behaviors observed in the experiments could result from the interplay between the distribution of contractile force and mechanosensitive bonds. 
\end{abstract}	
\maketitle 

{\it Introduction.--}
The crawling motion of eukaryotic cells is ubiquitous in biology as it plays important roles in processes such as embryogenesis, wound healing, cancer metastasis, and immunology~\cite{ref:Bray_book}.  
Common if not universal features of a crawling cell include myosin motors distributed mainly behind the center, dominant actin polymerization in the leading edge, and higher density of adhesion complexes in the leading region~\cite{ref:Theriot_JCB_2007}.  
Such polarized molecular distribution enables protrusion in the leading edge due to actin polymerization, treadmilling of actomyosin cytoskeleton due to contractility, and traction force pulling the cell body.    
These features, therefore, are included in many theoretical models for crawling cells~\cite{ref:Aranson_book}.

Interestingly, besides non-motile resting and steady-moving behaviors, cells crawling along a one-dimensional track either on a substrate or in a three-dimensional environment also exhibit moving patterns that are non-stationary in time.  
For example, stick-slip crawling motion due to slip between integrin and the extracellular matrix in focal adhesions under the contractility provided by myosin II has been observed in human osteosarcoma cells~\cite{ref:Gardel_CurrBiol_2010}.  
Periodic back-and-forth migration has been observed in crawling zyxin-depleted cells in a collagen matrix~\cite{ref:Wirtz_NatComm_2012} and dendritic cells crawling along microfabricated channels~\cite{ref:Chabaud_NatComm_2015}.  

Several theoretical models have been proposed to explain some of these deterministic complex moving patterns.
A model that includes the mechanochemical coupling of actin promotor dynamics and actin polymerization 
to myosin kinetics was shown to produce periodic back-and-forth migration~\cite{ref:Levine_PRL_2013}.  
On the other hand, a purely mechanical model emphasizing the interplay between mechanosensitive bonds and membrane tension exhibited stick-slip motion even for slip bonds~\cite{ref:Sens_PNAS_2020}.  
Interestingly, it has also been shown that stick-slip can result from the interplay between mechanosensitive bonds, contractility, and a force that tends to restore a cell's preferred length~\cite{ref:Gov_PRR_2020}.

In this {\em letter}, we present a theoretical study to show that the coupling between mechanosensitive adhesion complexes and myosin contractility is sufficient to generate deterministic complex cell crawling behaviors, 
including stick-slip, periodic back-and-forth movements, and other complex moving patterns.   
We first construct an active gel model with mechanosensitive adhesion complexes and show that  
the distribution of myosin motors and adhesion complexes computed from our model agree with experimentally 
observed features.  
By exploring the motility behavior with different strengths of contractility and mechanosensitivity, we show that this model can lead to complex motility behaviors other than rest and constant-velocity moving states.  
Among these complex motility patterns, unidirectional stick-slip motion and periodic back-and-forth movement are the most common.  
When the adhesion complexes are less mechanosensitive, as myosin contractility increases, the cell performs constant velocity motion.   
On the other hand, for cells with highly mechanosensitive adhesion complexes, as myosin contractility becomes sufficiently strong, periodic back-and-forth crawling motion can be observed.  
Finally, stick-slip and other complex motility patterns can be observed by increasing  
the mechanosensitivity of the adhesion complexes for a cell moving at constant velocity.  

To understand the physical mechanisms that produce these complex motility patterns, 
a simplified model inspired by the active gel model is constructed.
The simplified model assumes that 
the dynamics of a sufficiently slow-moving cell are dominated by 
the dipole moment of myosin density and the difference 
in the total number of adhesion complexes near the two cell ends.   
Remarkably,  the motility behavior predicted by this simple model agrees qualitatively with the motility behaviors predicted by the active gel model.  
These results suggest that, in general, diverse complex motility behaviors can result from the interplay between mechanosensitive adhesions and the dynamical organization of contractile myosin motors. 

{\it Model summary.--} 
To focus on the contribution of cell mechanics to motility behaviors, chemical signaling is not included in our model. 
The cytoplasm of the cell is 
modeled as an active gel~\cite{ref:Prost_2007,ref:Truskinovsky_PRL_2013} enclosed by the cell membrane, the adhesion complexes are treated as reversible bonds with specific binding-unbinding rates, and actin polymerization is assumed to happen only at the cell ends.  
The forces acting on the cell include the stress in the cytoskeleton, the drag force from the substrate, and the force due to the adhesion complexes.  

Our model only considers one spatial direction, {\it i.e.}, the cell's moving direction, and the stress in the cytoplasm obeys the constitutive equation
\begin{equation}
    \sigma = \eta \frac{\partial v}{\partial x} + \chi c ,
    \label{eq:stress}
\end{equation}
where $\eta$ is the effective one-dimensional viscosity of the cytoplasm, $v$ is the flow field, $\chi$ is the strength of contractility provided by myosin motors ($\chi >0$), and $c$ is the concentration of myosin attached to the actin network. 
For simplicity, compressibility is not included~\cite{ref:Truskinovsky_PRL_2013}. 
Thus pressure does not appear in the constitutive relation.  
The force exerted by the substrate is 
\begin{equation}
    F_{drag} =  - \alpha n_b v - \xi v,
    \label{eq:drag}
\end{equation}
The first term on the right hand is the drag provided by the adhesion complexes~\cite{ref:sekimoto1991}, 
$\alpha$ is a constant that characterizes the resistance of the adhesion complexes to cell movement, and $n_b$ is the number density of adhesion complexes. 
The second term comes from the viscous drag of the fluid between the cell and the substrate, and $\xi$ is the drag coefficient.  
Putting Eqs.~(\ref{eq:stress})(\ref{eq:drag}) together, the resulting force balance equation, $\partial _x \sigma  + F_{drag} = 0$, takes the following form
\begin{equation}
    \eta\frac{\partial^2 v}{\partial x^2} - (\alpha n_b  + \xi) v = -\chi \frac{\partial c}{\partial x}.\label{eq:1}
\end{equation}

Myosin motors attached to actin filaments move with the cytoplasm, while those detached from actin filaments diffuse freely.  
The attachment/detachment of motors is reversible.  On long-time scales, the density of the motors can be effectively described by an advection-diffusion equation \cite{ref:Recho_2016}
\begin{equation}
    \frac{\partial c}{\partial t} = D \frac{\partial^2 c}{\partial x^2} - \frac{\partial (c v)}{\partial x},
    \label{eq:2}
\end{equation}
where $D$ is the effective diffusion coefficient of myosin motors.

Adhesion complexes providing anchorage to the extracellular matrix are also physically coupled to the contractile cytoplasm.  As a result, they are pulled when the cytoplasm moves~\cite{ref:Sens_PNAS_2020}.  
Once an adhesion complex is formed, the adhesion site does not move, but the dissociation rate of the adhesion complex is affected by the motion of the cytoskeleton because the bond is stretched or compressed.  
In our model, the evolution of the density of adhesion complexes is assumed to obey
\begin{equation}
    \frac{\partial n_b}{\partial t} 
    = -k_0 \ e^{-k_1 \partial _x v} n_b + k_{\rm on},
    \label{eq:nb}
\end{equation}
where $k_{\rm on}$ is the binding rate, $k_0$ is the unbinding rate at $\partial _xv=0$, and $k_1$ tells us how unbinding rate is affected by the cytoplasmic flow.  
Our model assumes that when the strain rate is dilating, giving more space for the adhesion complexes, the unbinding rate decreases.  

Actin polymerization at the cell ends depends on the distribution of actin activators~\cite{ref:Firtel_Cell_2003}\cite{maeda2008ordered}.  
In the presence of environmental cues, a gradient of actin activator concentration within the cell is established, and actin polymerization is polarized due to this concentration gradient.  
In the absence of such external influence, the cell can nevertheless polarize itself by spontaneous symmetry breaking, and the net actin polymerization rate becomes asymmetric. 
In our model, we consider a homogeneous environment, and the net actin polymerization rate $v_p^{\pm}$ at the $\pm$ end of the cell is assumed to be   
\begin{equation}
    v_p^{\pm} = \frac{2 \ e^{-v_p^{(1)}(L-L_0)}}{1+\exp [\mp \frac{dl_{\pm} }{dt}/v_p^{(2)}]}v_p^{(0)},
    \label{eq:vp}
\end{equation}
where $v_p^{+}$ ($v_p^{-}$) is the net rate of extension due to actin polymerization at the cell end located at $x = l_{+}(l_{-})$,  $v_p^{(0)}$ comes from the base polymerization rate, 
$v_p^{(1)}$ in the exponent of the numerator comes from the effect of free energy cost for polymerization when the cell length is different from its natural length ($L_0$ is the natural length of the cell, and $L = l_{+}-l_{-}$), 
and the term with $v_p^{(2)}$ makes the net polymerization rate in a moving cell at both ends different, with more polymerization events in the leading end than the trailing end.  
It will become clear that the qualitative results of cell motility behavior do not depend on the specific form we assumed for the dissociation rate of the adhesion complexes and $v_p^{\pm}$. 

The evolution of cell-end positions is determined by the velocity of cytoplasm and actin polymerization, 
\begin{equation}
        \frac{d l_{\pm}}{d t} = v_{\pm} \pm v_p^{\pm},\label{eq:4}
\end{equation}
where $v_{+}$ ($v_{-}$) is the velocity of cytoplasm at the $+ \ (-)$ end.

Experimentally it has been shown that a cell tends to restore its length $L$ to a preferred magnitude $L_0$~\cite{ref:Tsai_DevCell_2019}.  We model this effect by the following force balance condition at cell ends 
\begin{equation}
        \sigma_{\pm} = \left[\chi c + \eta \frac{\partial v}{\partial x} \right]_{l_{\pm}} = - \gamma (L - L_0).
\label{eq:cell_end_stress}
\end{equation}
Here $\gamma$ is a constant associated with the restoring force that brings the cell length $L$ to $L_0$. 

Because no myosin motors can leave or enter the cell, the total flux of myosin motors across a cell end should vanish.  
This leads to
\begin{equation}
        \left[c v \right]_{l_{\pm}} - \left[c\right]_{l_{\pm}} \frac{dl_{\pm}}{dt} - D \left[\frac{\partial c}{\partial x} \right]_{l_{\pm}} = 0.\label{eq:7}
\end{equation}
The first two terms on the left-hand side are the advective flux relative to the moving cell end, and the last term is the diffusive flux at the cell end. 

It is convenient to introduce effective drag coefficient $\xi _{\rm eff} = \xi + \alpha k_{\rm on}/k_{0}$ and  choose $l_0 = \sqrt{\eta / \xi _{\rm eff}}$ as the unit length, 
$t_0 =\eta / (\xi _{\rm eff} D)$ as the unit time, $\sigma_0 = \xi _{\rm eff} D$ as the unit stress, 
$n_0 = \xi /\alpha$
as the unit density for adhesion complexes, 
and $c_0 = M/ \sqrt{\eta / \xi _{\rm eff}}$ as the unit myosin concentration, where $M$ is the total number of myosin motors in the cell.
Therefore 
the dimensionless drag coefficient $\tilde{\xi } = \xi / (\xi +\alpha k_{\rm on}/k_{0})$,  
contractility $\tilde {\chi} = c_0 \chi / \sigma _0$, and cell elastic constant $K = \gamma l_0 /\sigma _0$
are used in the following discussion.

{\it Simulation of motility behaviors.--}
From the point of view of nonequilibrium thermodynamics, the drag force between the cell and the substrate, the viscous force in the cytoplasm, and the diffusion of myosin are passive processes against cell movement.
On the other hand, active processes such as actin polymerization and myosin contractility drive the movement of the cell, and the binding/unbinding dynamics of the adhesion complexes modulate cell movement. 
As the myosin motors provide contractility against viscous and substrate drag, the contractility-induced cytoplasmic flow drifts the motors to aggregate and also affects the distribution of adhesion complexes.  
Once the cell is in motion, the feedback in the actin polymerization rate further enhances cell polarization.  
The balance between these processes determines the state of the cell. 
In general, there is no analytical solution when all these effects are included.  Therefore we numerically integrate the equations of motion by a finite difference method.      
The details of our numerical methods and our choice of parameters are discussed in~\cite{ref:SI}.  

\begin{figure}[tbh]
\mbox{
\subfigure[]{\includegraphics[width=.95\linewidth]{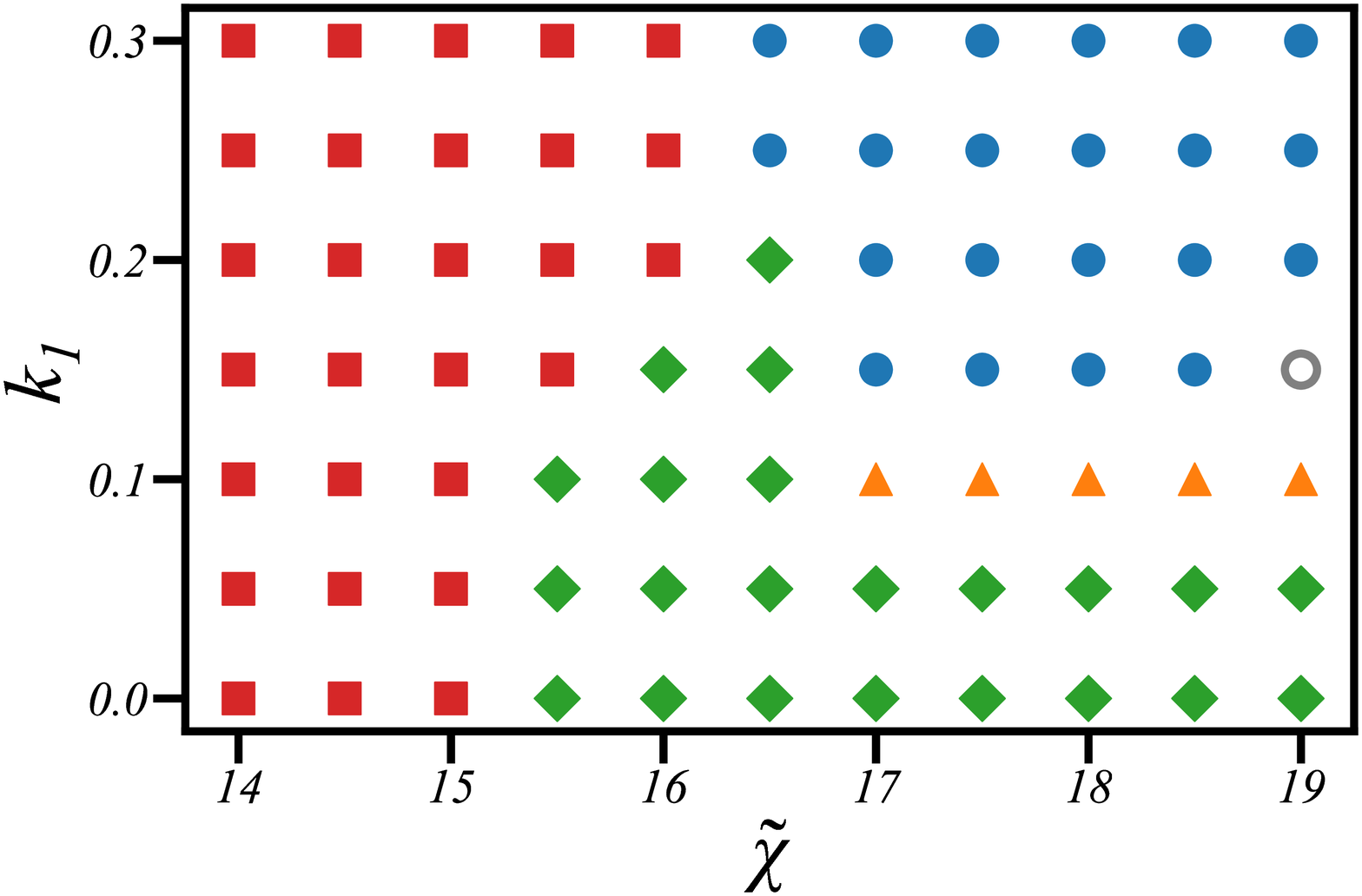}}
}
\mbox{
\subfigure[]{\includegraphics[width=.25\linewidth]{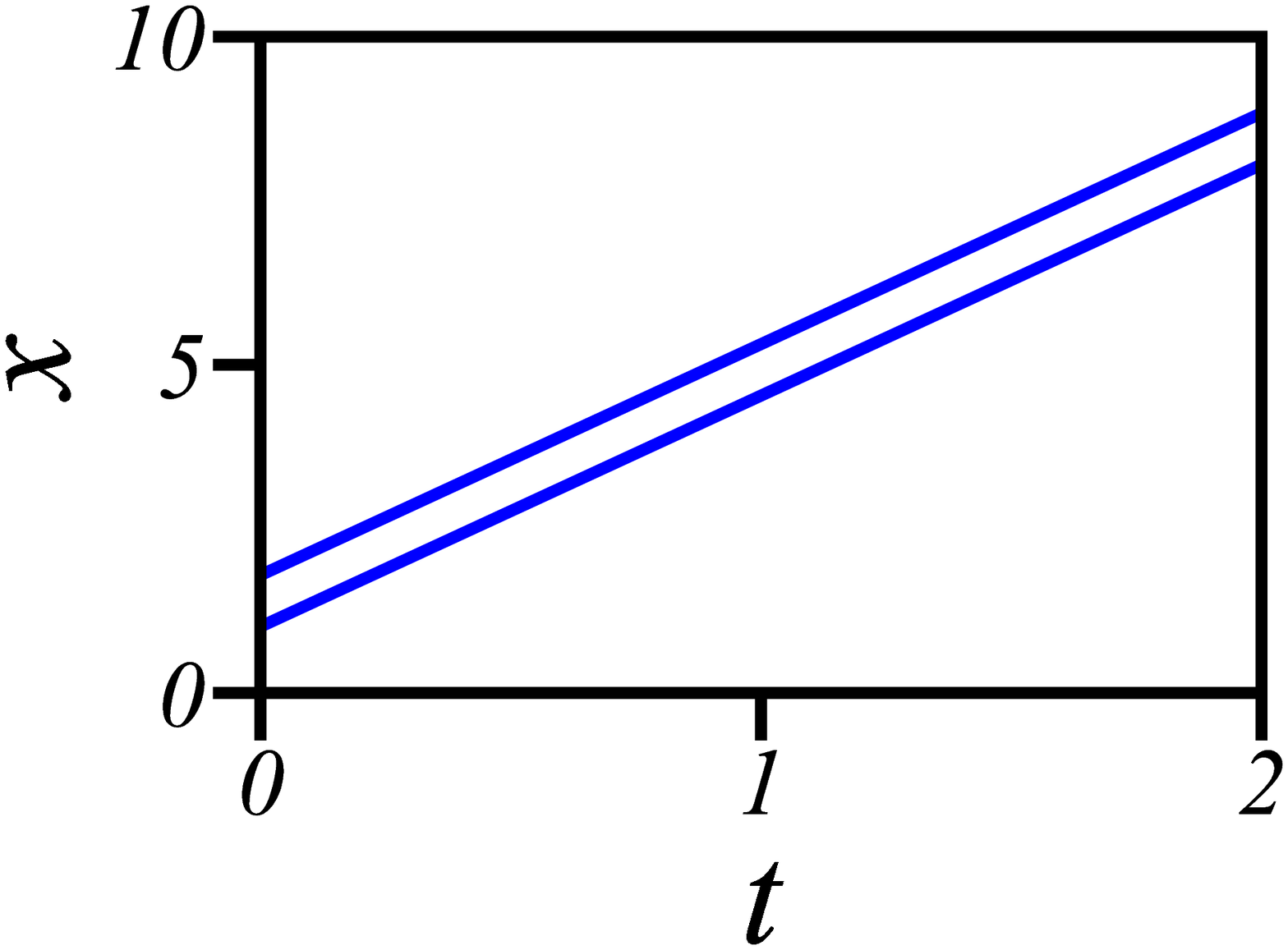}}
\subfigure[]{\includegraphics[width=.25\linewidth]{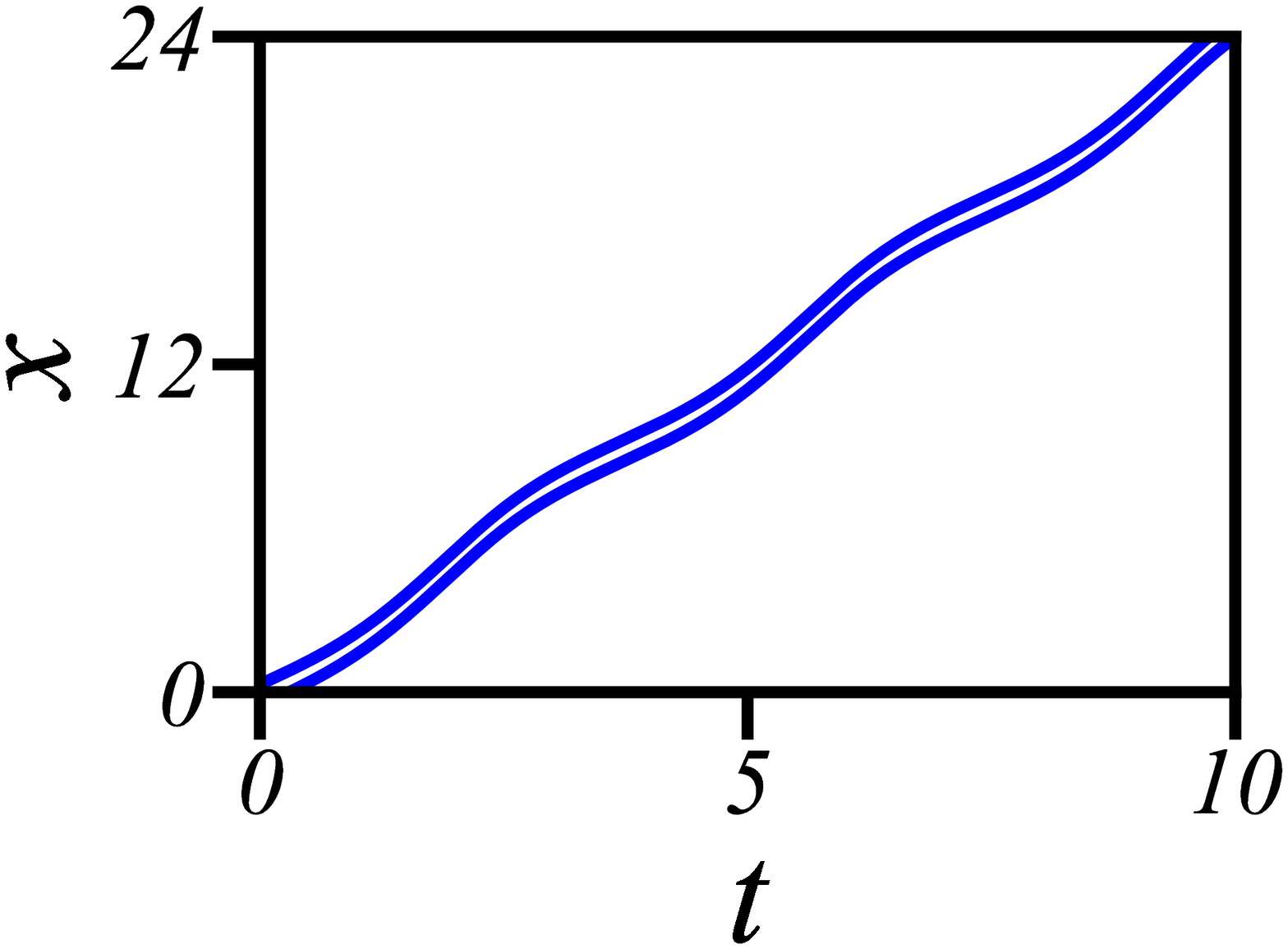}}
\subfigure[]{\includegraphics[width=.25\linewidth]{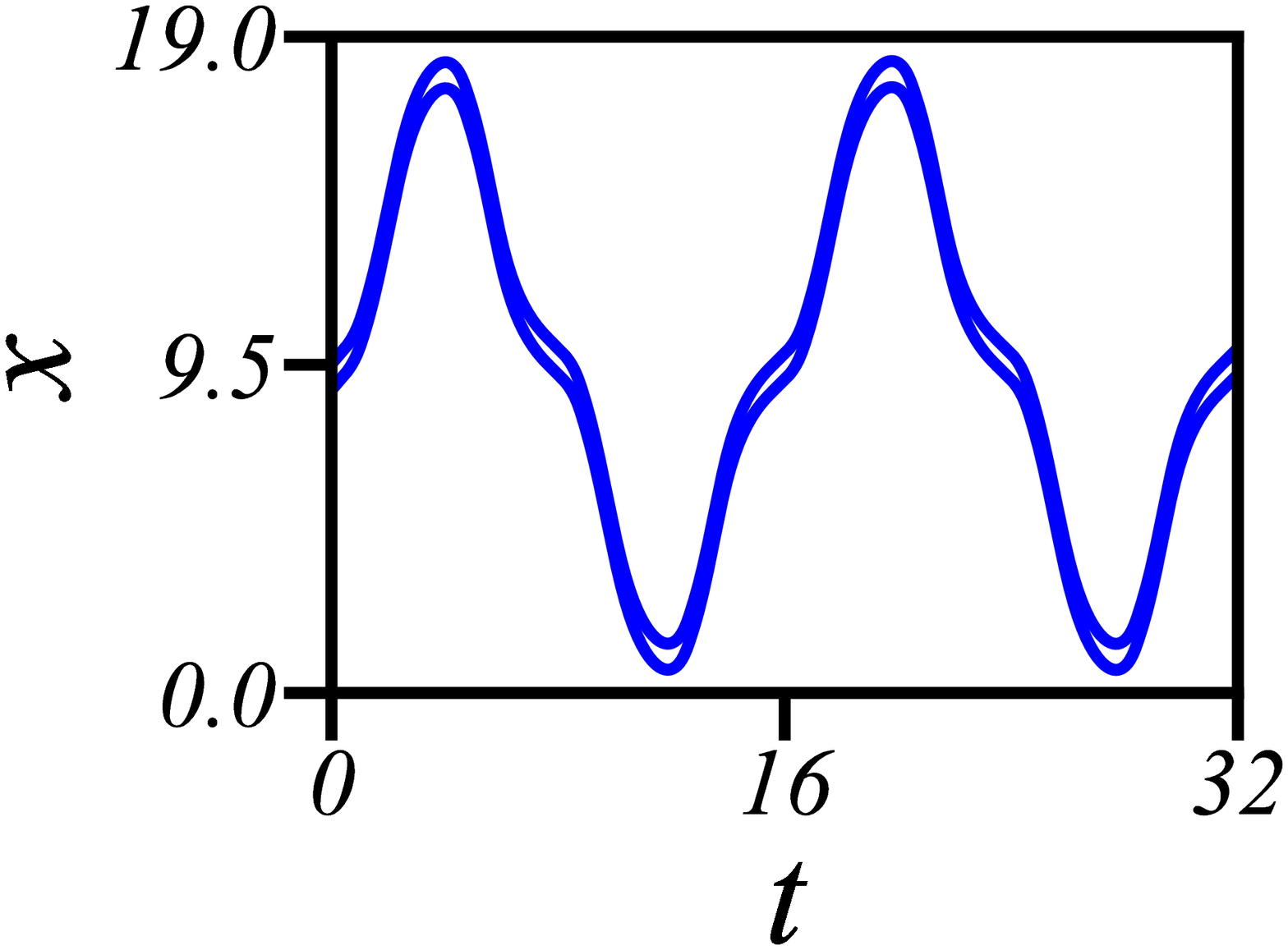}}
\subfigure[]{\includegraphics[width=.25\linewidth]{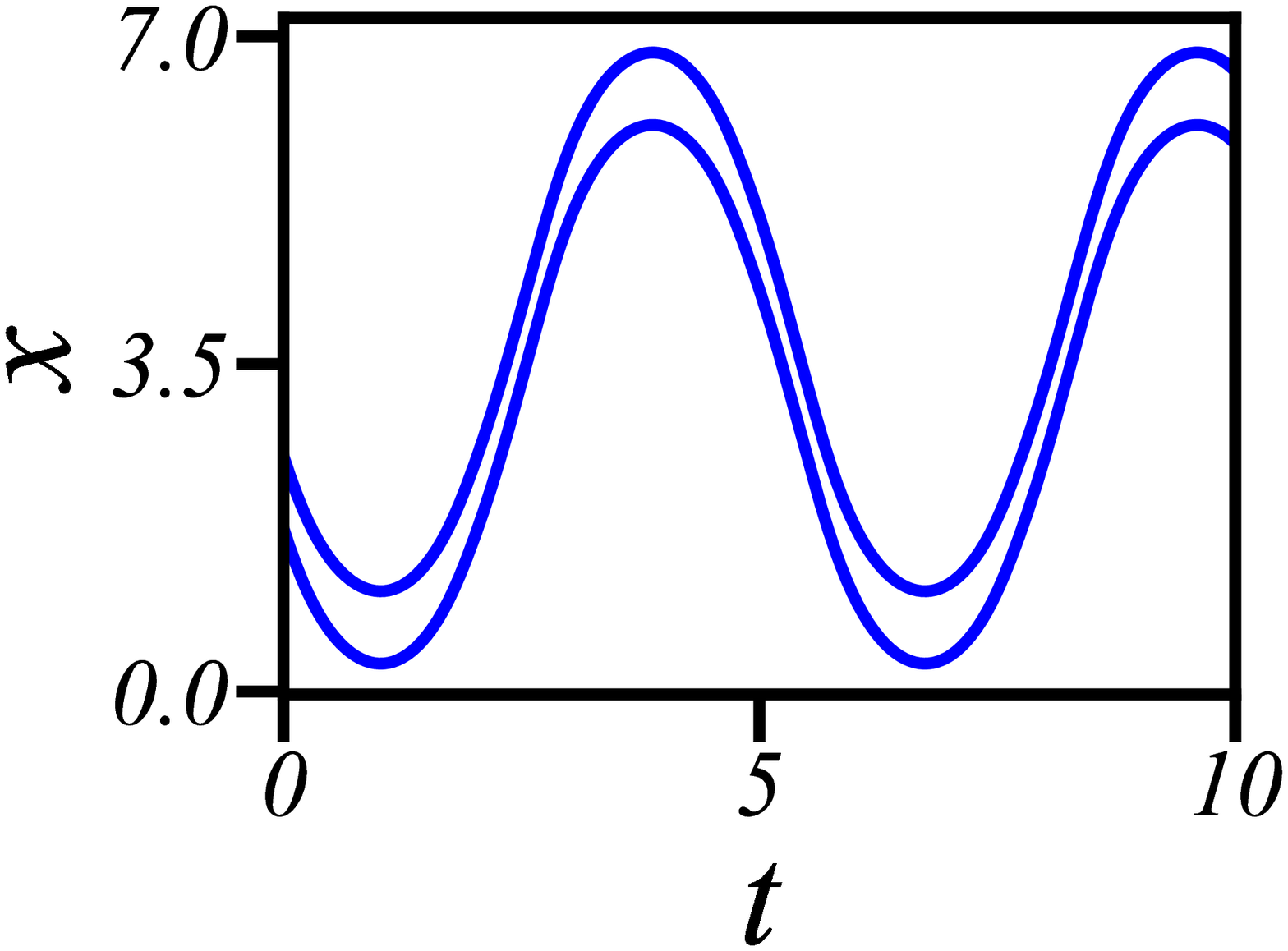}}
}
\caption{ (a) Motility phase diagram for a cell with dimensionless parameters $K = 100$, $\tilde{\xi} = 1/3$, $k_{\rm on}= 6$, $k_0=3$, $v_p^{(0)}=0.2$, $v_p^{(1)}=0.5$, and $v_p^{(2)}=2$.  
Rest state (squares), constant-velocity motion (diamonds), stick-slip movement (triangles), 
back-and-forth with stick-slip motion (empty circle), and periodic back-and-forth motion (filled circles) are found. 
(b) Trajectories of the cell ends for $\tilde{\chi} = 18$, $k_1 = 0.05$, the cell performs constant velocity motion. 
(c) Trajectories of the cell ends for $\tilde{\chi} = 17.5$, $k_1 = 0.1$, the cell performs stick-slip motion.
(d) Trajectories of the cell ends for $\tilde{\chi} = 19$, $k_1 = 0.15$, the cell performs complex motility pattern 
which is periodic back-and-forth with stick-slip.
(e) Trajectories of the cell ends for $\tilde{\chi} = 18$, $k_1 = 0.25$, the cell performs periodic back-and-forth motion.
}
\label{fig:phase_diagram}
\end{figure}

Figure~\ref{fig:phase_diagram} shows the motility behaviors for a cell with parameters chosen to be compatible with typical cells~\cite{ref:SI} and a range of adhesion complex mechanosensitivity and contractility strengths.
The following motility behaviors are found: rest, moving at a constant velocity, unidirectional stick-slip movement, back-and-forth motion with stick-slip, and periodic back-and-forth movement. 
For a cell with weakly mechanosensitive adhesion complexes, as contractility increases, a cell at rest starts to move at constant velocity.
As the adhesion complexes become more mechanosensitive, a moving cell shows other complex motility behaviors.  
For example, stick-slip motion and (at high contractivity) back-and-forth motion with stick-slip, and finally, the cell performs periodic back-and-forth motion when the adhesion complexes are highly mechanosensitive.  
Another motility phase diagram in the Supplement Materials~\cite{ref:SI} shows that, within our model,  a cell 
with a high actin polymerization rate can exhibit other complex motility behaviors between stick-slip and periodic 
back-and-forth movements. 

\begin{figure}[tbh]
\begin{center}
\mbox{
\subfigure[]{\includegraphics[width=.45\linewidth]{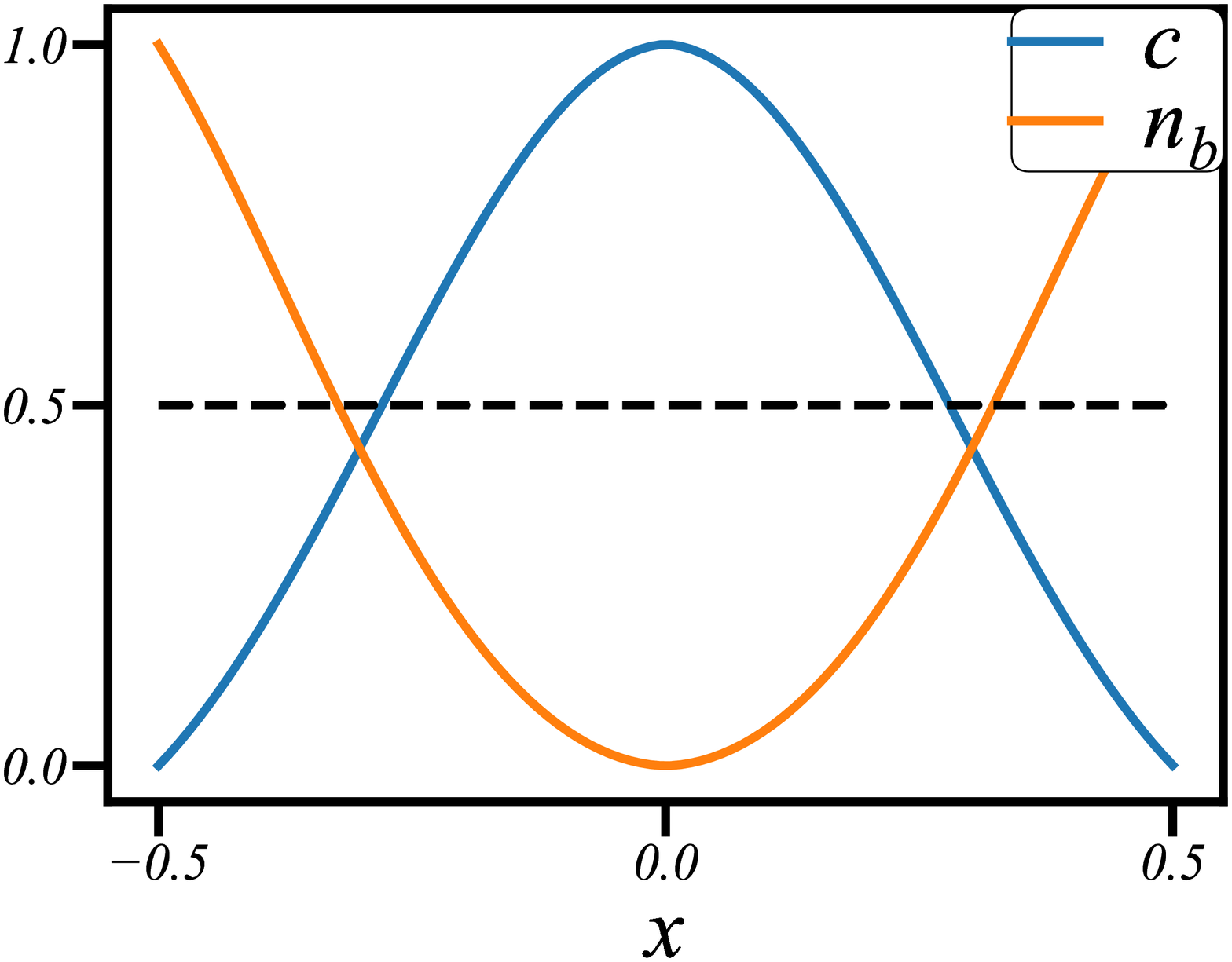}}
}
\mbox{
\subfigure[]{\includegraphics[width=.45\linewidth]{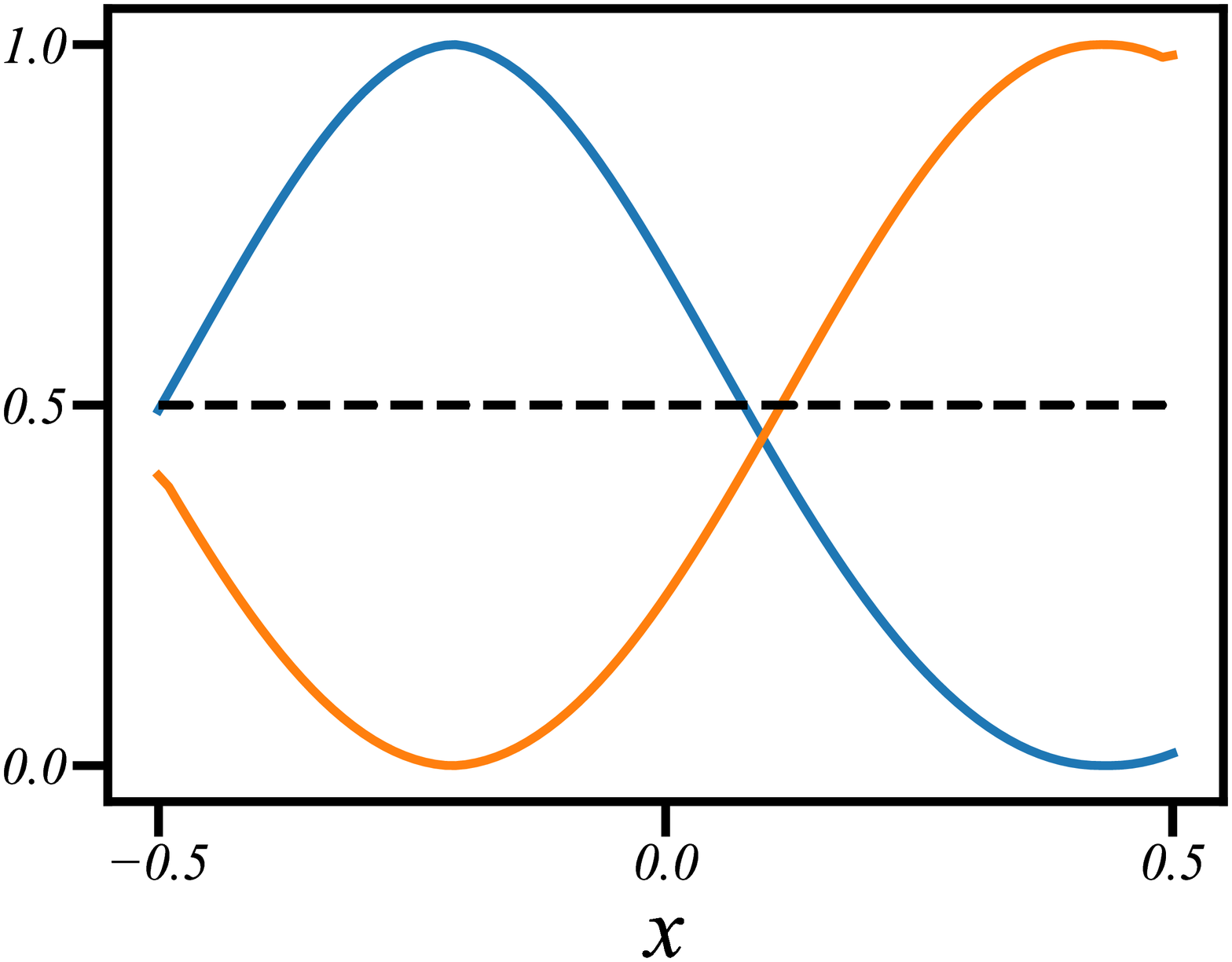}}
}
\end{center}
\caption{ 
Distribution of myosin motors and adhesion complexes for a cell (a) at rest and (b) undergoes constant velocity motion towards $+x$ direction.  The parameters are the same as those in Fig.~1 and (a) $k_1 = 0.25$, $\tilde{\chi} = 16$, 
(b) $k_1 = 0.1$, $\tilde{\chi} = 16$. }
\label{fig:distributions}
\end{figure}

Figure~\ref{fig:distributions} shows that when the cell is at rest, myosin motor distribution is symmetric around the center of the cell, and the number of adhesion complexes near both cell ends is the same;
on the other hand, for a cell moving at constant velocity, myosin motors aggregate close to the trailing end and 
adhesion complexes are mainly close to the leading end.  
The distribution of myosin motors and adhesion complexes  
for a cell undergoes stick-slip, and periodic back-and-forth movements are shown in Fig.~S2 of~\cite{ref:SI}. 
It is clear that whenever the cell has a definite moving direction, myosin motors aggregate in a regime behind the 
center of the cell, and more adhesion complexes form near the leading end than the trailing end.  
Indeed, the adhesion complex binding/unbinding rates~Eq.(\ref{eq:nb}) and net actin polymerization at the cell ends~Eq.(\ref{eq:vp}) in our model lead to reasonable molecular distributions in a cell.  

{\it Reduction to the simplified model.--}
To obtain an intuitive physical picture of the complex motility behaviors, especially the transitions from the rest state to the constant-velocity movement and periodic back-and-forth movement, 
we construct a simplified model from the active gel model.  
First, we consider a limiting situation in which adhesion complexes only appear in a small region close to the cell ends.   
In this regime, it is convenient to introduce $N_f (N_b)$, the total number of adhesion complexes close to $l_+(l_-)$, and $N= N_f +N_b$, $\Delta N = N_f - N_b$ to describe the distribution of adhesion complexes.  
We also introduce $y_c$, the dipole moment of myosin motors density relative to the center of the cell~\cite{ref:SI}, 
to characterize the spatial distribution of myosin motors.  
The net actin polymerization velocity at the cell ends $v_p^{\pm} \equiv v_p \pm \Delta v_p /2$,
where  $\Delta v_p = v_p^{+} - v_p^{-} \propto V_{\rm cell}$ in the limit of small cell velocity. 
Therefore we write  $v_p^{\pm} = v_p \pm \beta V_{\rm cell} /2$. 
In this regime, straightforward calculation shows that the velocity of the cell is ~\cite{ref:SI} 
\begin{eqnarray}
V_{\rm cell} \approx \frac{1}{1-\beta/2}(\lambda _{\nu 1} v_p \Delta N - \tilde{\chi} \lambda_{\nu 2} y_c),
\end{eqnarray}
where $\lambda _{\nu 1}$ and $\lambda _{\nu 2}$ are positivet coefficients that depend on $N$ and $L$. 
Note that, from the definition, $\beta/2$ cannot be greater than unity.  Therefore a cell with $\Delta N > 0$ and $y_c < 0$ has positive $V_{\rm cell}$.  
This is in agreement with experimental observations.

We further simplify the analysis by considering the limit of large $K$, {\it i.e.}, $L \approx L_0$.  
The following equations are constructed to describe the dynamics of $N$,  $\Delta N$, and $y_c$. 
First, the evolution equations for $N$ and $\Delta N$ are
\begin{eqnarray}
\frac{dN}{dt} &=&  2 k_{\rm on} - k_{\rm off} ^{(0)} \ N - k_{\rm off}^{(1)} \ y_c \ \Delta N  , \nonumber \\ 
\frac{d\Delta N}{dt} &=& - k_{\rm off}^{(0)} \Delta N - k_{\rm off}^{(1)} \ N y_c,
\label{eq:dNdtdDeltaNdt}
\end{eqnarray}
where $k_{\rm on}$, $k_{\rm off}^{(0)}$, and $k_{\rm off}^{(1)}$ are positive constants.  
Next, in the spirit of Landau-type approximation, the evolution of $y_c$ is assumed to obey the following 
equation, 
\begin{eqnarray}
\frac{dy_c}{dt} = - \Gamma \left[ - (\tilde{\chi} - \tilde{\chi}_c ) y_c -  a_{\Delta N} \Delta N + a_3 y_c^3 \right],
\label{eq:dycdt}
\end{eqnarray}
where $\tilde{\chi}_c$ and $a_{\Delta N}$ 
are treated as $N$-independent constants for simplicity.  

The simple model equations~(\ref{eq:dNdtdDeltaNdt})(\ref{eq:dycdt}) have a solution with constant $N$ and 
$\Delta N = y_c = 0$.  This corresponds to a cell at rest. 
Solutions with nonzero constant $\Delta N$ and $y_c$ correspond to a cell moving at a constant velocity; 
solutions with time-periodic $\Delta N$ and $y_c$ are stick-slip (periodic back-and-forth) movement 
if the time-average of $\Delta N$ and $y_c$ are nonzero (zero).
The linear stability analysis of the rest state shows that as the contractility increases, 
a cell at rest starts to move as the system undergoes a bifurcation, 
the moving state is the constant-velocity state 
when $k_{\rm off}^{(1)}$ is small, i.e., the adhesion complexes are less mechanosensitive.  
When $k_{\rm off}^{(1)}$ is sufficiently large, the rest-to-moving transition leads to 
a periodic back-and-forth moving state~\cite{ref:SI}.

As shown in Fig.~S3~\cite{ref:SI}, the model Eqs.~(\ref{eq:dNdtdDeltaNdt})~(\ref{eq:dycdt}) exhibit a motility phase diagram qualitatively the same as the numerical solutions of our active gel model. 
The minor differences come from those simplifications made when constructing the simplified model.  
Furthermore, 
from the simplified model, it is easy to see how the symmetry properties and the couplings of the key driving variables lead to the observed cell motion.  
For example, $y_c(t)$ and $\Delta N(t)$ in Fig.~3(a) for a cell performing periodic back-and-forth movement suggests the following physical picture about how the coupling terms in the simplified model lead to this motion.  
According to Eq.~(\ref{eq:dNdtdDeltaNdt}), a small $\Delta N$ tend to increase when $y_c$ is sufficiently negative, and Eq.~(\ref{eq:dycdt}) states that when $a_{\Delta N} >0$,  $y_c$ tends to move toward the center of the cell 
when the magnitude of $\Delta N$ is sufficiently large. 
The result is that at sufficiently large $k_{\rm off}^{(1)}$, the number of adhesion complexes in the leading end of a moving cell increases sufficiently fast such that at some point, the myosins are pulled to the other half of the cell, reversing the sign of $y_c$, then reversing the sign of $\Delta N$,  and eventually the direction of cell motion is reversed.
This is how rest/constant-velocity transition becomes rest/back-and-forth transition as the adhesion complexes are more mechanosensitive.  
In between the constant velocity and periodic back-and-forth movement, complex motility patterns with stick-slip can be observed. 
As shown in Fig.~3(b), $y_c(t)$ and $\Delta N(t)$ 
in a cell that undergoes stick-slip movement have nonzero time-average values, and they oscillate with similar phase-relations as a cell undergoes periodic back-and-forth movement.  
This is because the mechanosensitivity of the adhesion complexes is sufficiently strong to induce an 
oscillation of $\Delta N$ and $y_c$, but not sufficiently strong to change their signs. 
Figure~3(c) and Fig.~3(d) show that $y_c(t)$ and $\Delta N(t)$ (defined as the difference of the total number of adhesion complexes in the leading and trailing halves of the cell) in the numerical simulations of the active gel model behave 
similarly, suggesting that the physical picture obtained from studying the simplified model can be applied to more detailed models.  
\begin{figure}[tbh]
\begin{center}
\mbox{
\subfigure[]{\includegraphics[width=.45\linewidth]{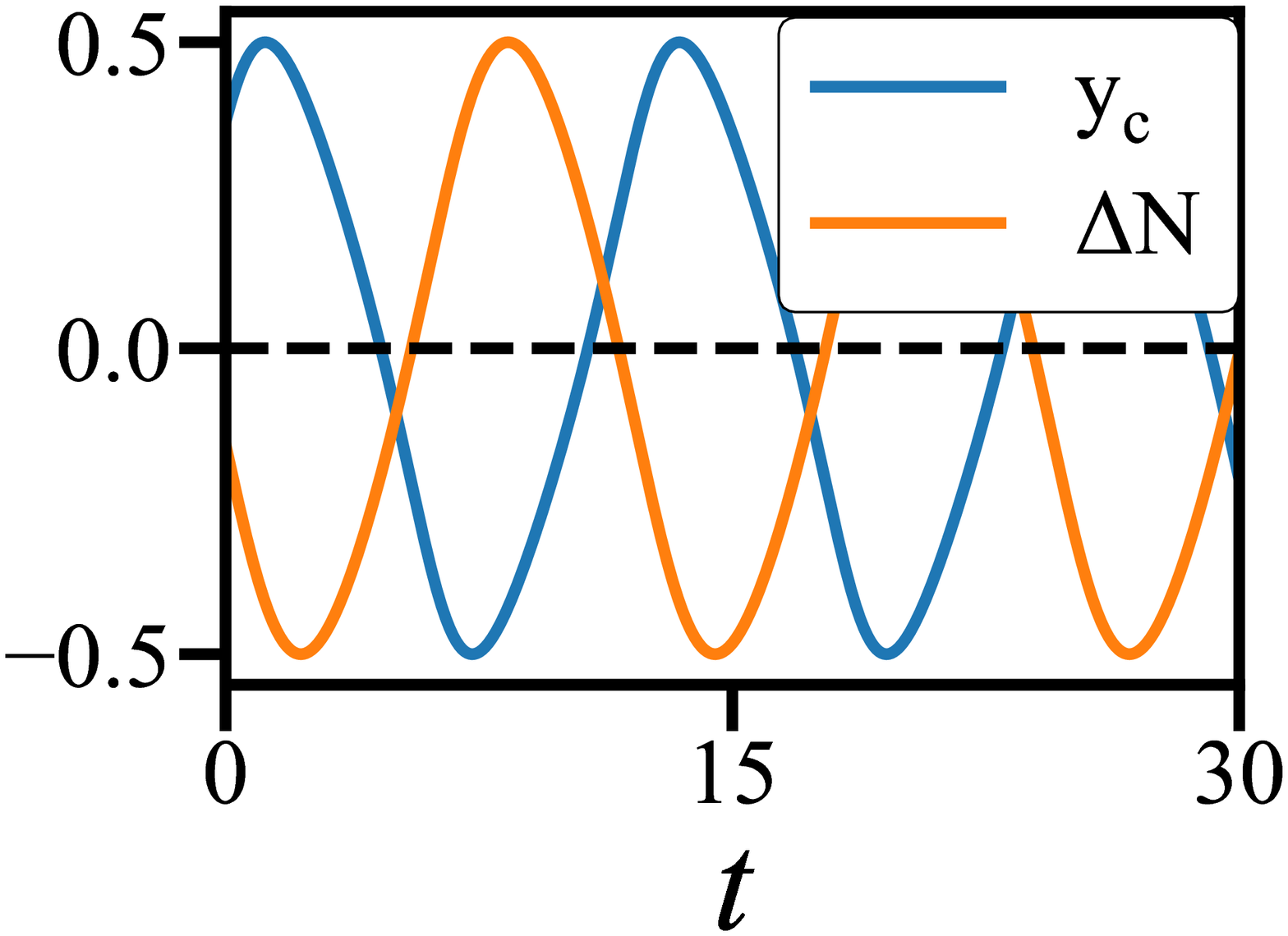}}
}
\mbox{
\subfigure[]{\includegraphics[width=.45\linewidth]{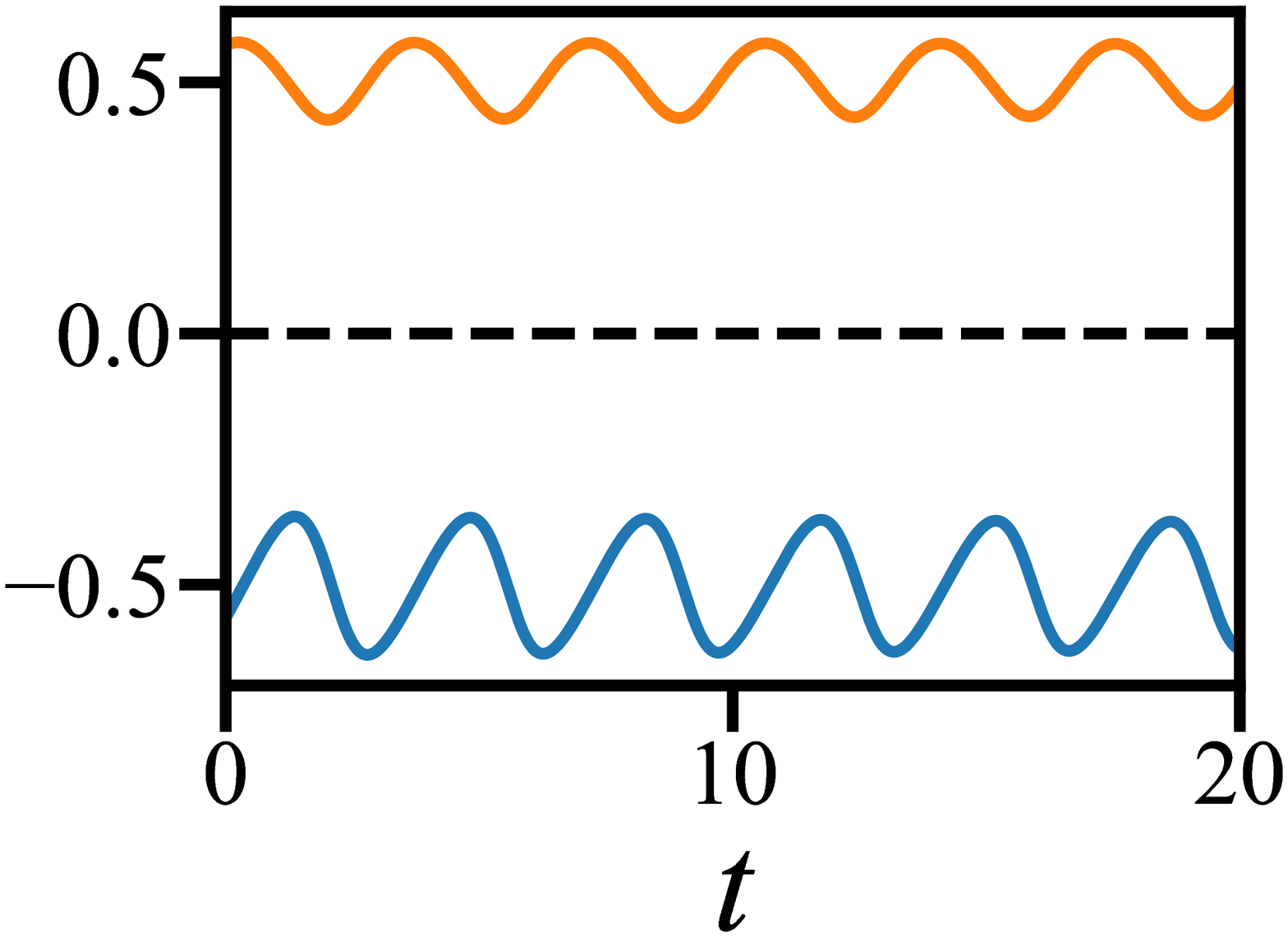}}
}
\mbox{
\subfigure[]{\includegraphics[width=.45\linewidth]{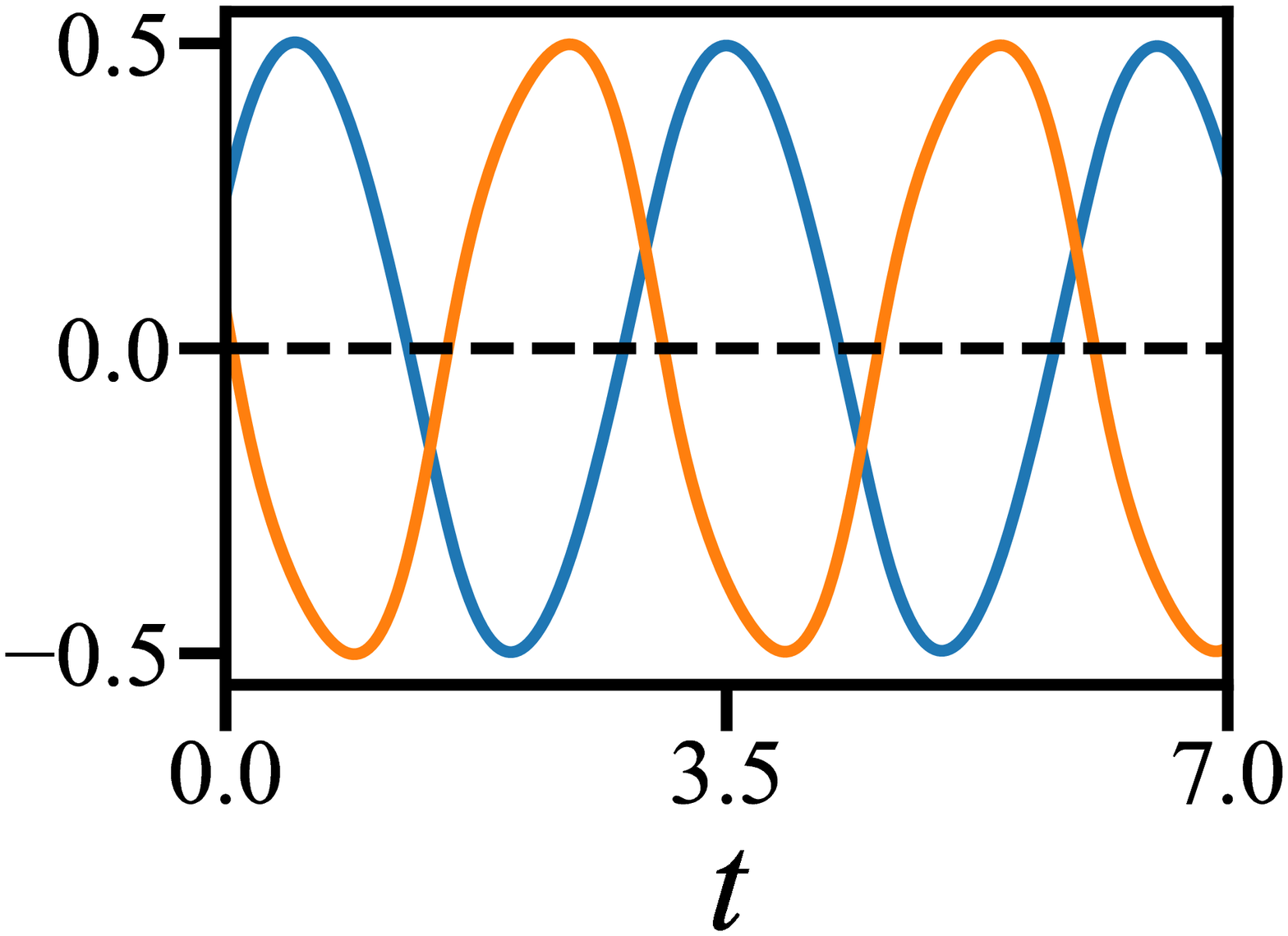}}
}
\mbox{
\subfigure[]{\includegraphics[width=.45\linewidth]{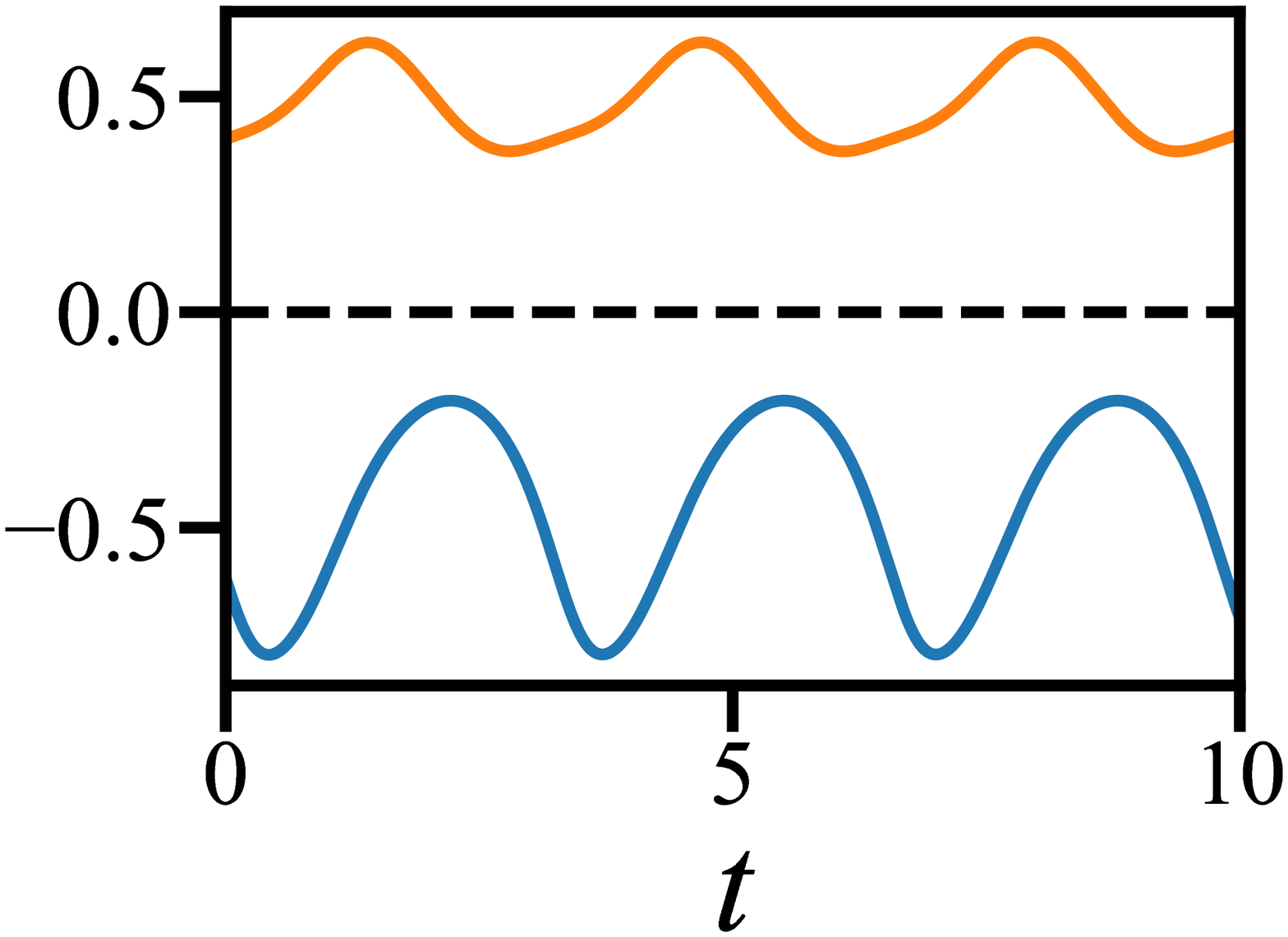}}
}
\end{center}
\caption{  (a)(b): $y_c(t)$ and $\Delta N(t)$ in the simplified model with $k_{\rm on}/k_{\rm off}^{(0)}=1$, $a_{\Delta N} = 1$, and $a_3 = 1$.
In (a),  $k_{\rm off}^{(1)}/k_{\rm off}^{(0)}=0.6$, $\Gamma (\tilde{\chi} - \tilde{\chi}_c)/k_{\rm off}^{(0)} = 1.2$; 
in (b),  $k_{\rm off}^{(1)}/k_{\rm off}^{(0)}=0.62$, $\Gamma (\tilde{\chi} - \tilde{\chi}_c)/k_{\rm off}^{(0)} = 5.5$.  
(c)(d): $y_c(t)$ and $\Delta N(t)$ in the active gel model with $K = 100$, $\tilde{\xi} = 1/3$, $k_{\rm on}= 6$, $k_0=3$, $v_p^{(0)} = 0.2$, $v_p^{(1)}=0.5$, $v_p^{(2)}=2$.
In (c), $k_1 = 0.55$, $\tilde{\chi}=16$; in (d),  $k_1 = 0.1$, $\tilde{\chi}=18$.
The cell in (a)(c) performs oscillatory back-and-forth movement, and the cell in (b)(d) performs stick-slip movement.
}
\label{fig:ycDN}
\end{figure}

{\it Discussion.--}  
Within our active gel model, a cell with highly mechanosensitive adhesion complexes can exhibit periodic back-and-forth movement similar to what was observed in zyxin-depleted cells in a collagen matrix. 
Since zyxin proteins act as mechanosensors in mature adhesion complexes~\cite{ref:Ingber_JCPhys_2006}, our study suggests that the difference in the mechanosensitivity of the adhesion complexes in zyxin-depleted and wild-type cells 
could be the origin of the periodic back-and-forth movement observed in~\cite{ref:Wirtz_NatComm_2012}. 
Future experiments can be designed to examine this prediction.  

The physical picture suggested by our simplified model also implies the possibility that, in general, when some of the simplifications are lifted, more complex one-dimensional cell motility behaviors can be found.  
This is indeed the case, as we explore the behavior of our active gel model for a broader range of actin polymerization rates, complex trajectories which come from further bifurcations are found. 
This is shown in Fig.~S1 in \cite{ref:SI} and Fig.~4.  
Further study of the physical mechanisms for these behaviors will be our future work~\cite{ref:future}. 

Finally, although the physical mechanisms for symmetry-breaking transitions, such as rest/periodic back-and-forth transition and rest/constant velocity transition, can be understood from the dynamics of $y_c$ and $\Delta N$, 
it is interesting to study how other important physical observables, such as the multipoles of the traction force~\cite{ref:Sano_BPJ}\cite{ref:Ohta_PhysicaD}, behave in cells with different moving patterns.  
It is also important to check if the basic features of these physical observables in different moving patterns depend on the details of the binding/unbinding dynamics of adhesion complexes, 
as it plays a significant role in our understanding of many interesting features of cell motility. 
\begin{figure}[tbh]
\begin{center}
\mbox{
\subfigure[]{\includegraphics[width=.4\linewidth]{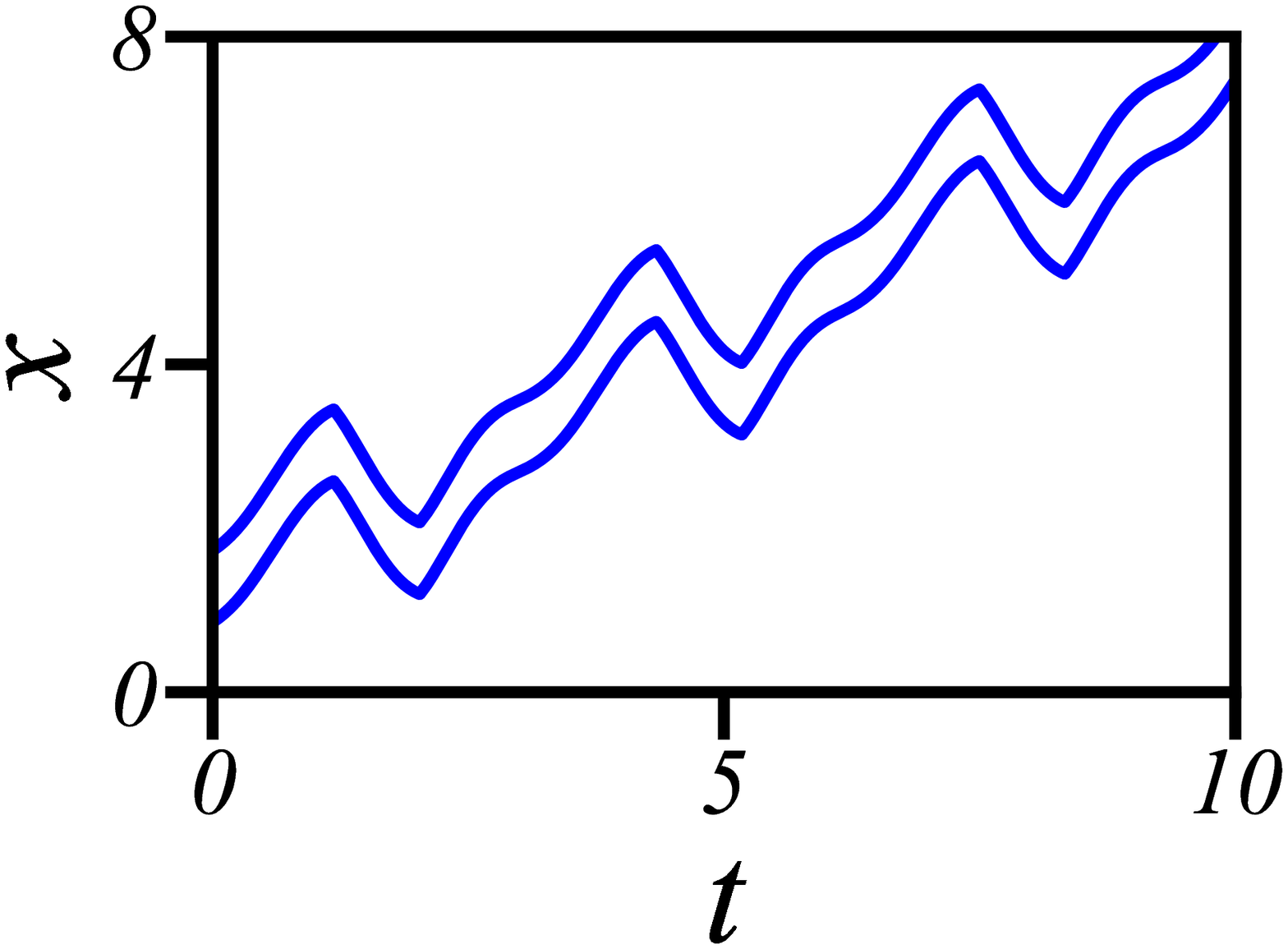}}
}
\mbox{
\subfigure[]{\includegraphics[width=.4\linewidth]{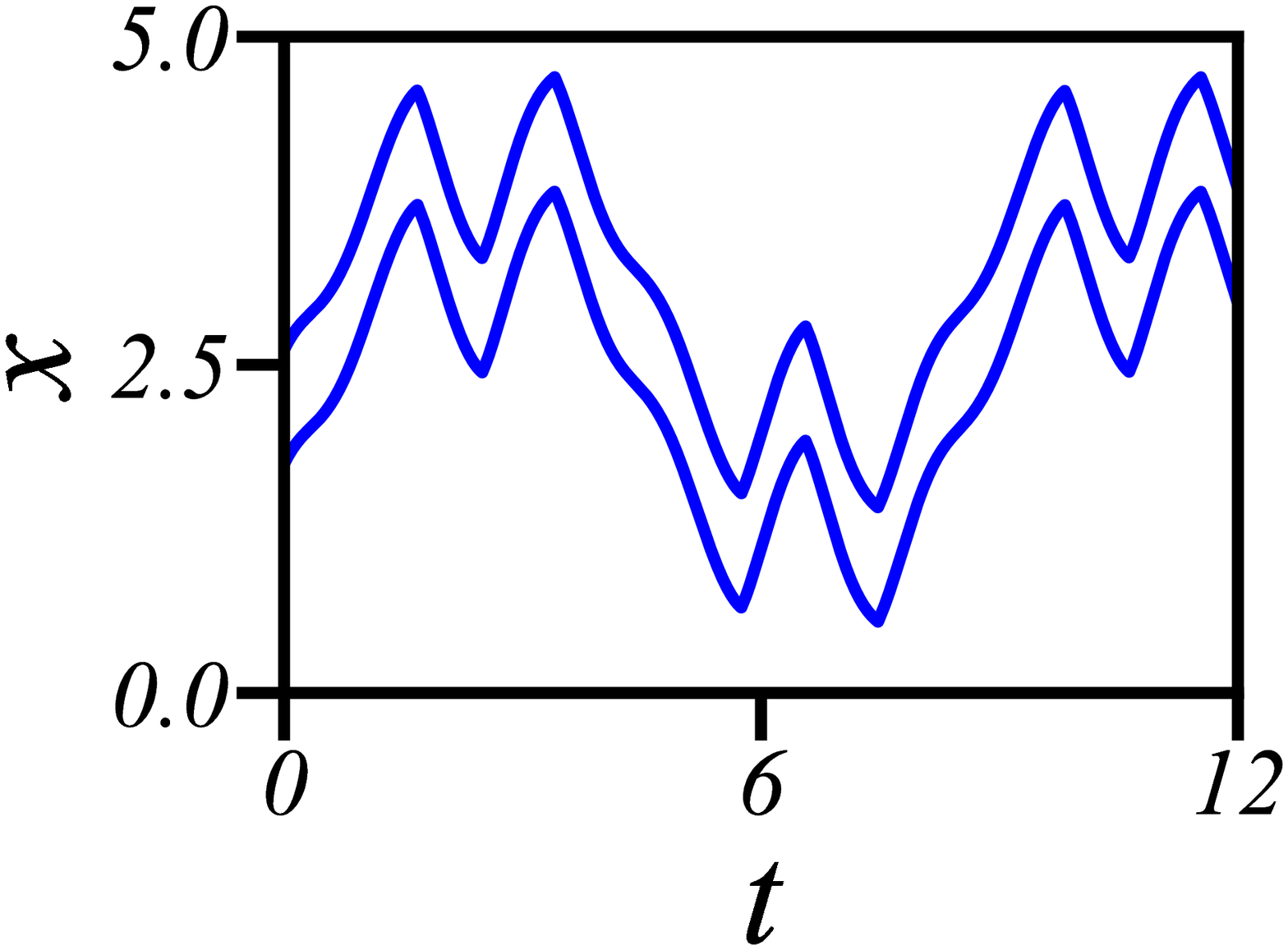}}
}
\end{center}
\caption{ 
Our active gel model predicts that at high actin polymerization rates further complex motility behaviors, such as 
(a) zig-zag with stick-slip, and 
(b) double-period back-and-forth movement can be found.    
These trajectorjes are obtained for $K = 100$, $\tilde{\chi} = 14$, $k_{\rm on} = 6.0$, $k_0 = 3.0$, $v_p^{(1)} = 0.5$, $v_p^{(2)} = 0.2$.
In (a), 
$v_p^{(0)} = 1.2$, $k_1=0.9$; in 
(b), $v_p^{(0)} = 1.1$, $k_1=0.9$.
The blue curves represent the trajectories of the cell ends.  
}
\label{fig:complex_phasediagram}
\end{figure}

{\it Acknowledgments --}

H.-Y. C. thanks Prof. Jasnow (University of Pittsburgh) for stimulating discussions and encouragement in the early stage of this work.
H.-Y. C. is supported by the Ministry of Science and Technology, Taiwan (MOST 108-2112-M-008-016 ). The authors also acknowledge the support from National Center for Theoretical Sciences, Taiwan.


\begin{thebibliography}{150}

\bibitem{ref:Bray_book}
D. Bray, 
{\it Cell Movements}, 2nd ed. 
(Talyor and Francis, New York, 2001).

\bibitem{ref:Theriot_JCB_2007}
P.T. Yam, C.A. Wilson, L. Ji, B. Hebert, E.L. Barnhart, N.A. Dye, P.W. Wiseman, G. Danuser, and J.A. Theriot,
 Actin–myosin network reorganization breaks symmetry at the cell rear to spontaneously initiate polarized cell motility,
 J. Cell Biol., {\bf 178}, 1207 (2007). 
 
 \bibitem{ref:Aranson_book}
I.S. Aranson ed., 
{\em  Physical Models of Cell Motility}, 
Springer, 2016.
 
 \bibitem{ref:Gardel_CurrBiol_2010}
Y. Aratyn-Schaus and M. L. Gardel, 
Transient frictional slip
between integrin and the ECM in focal adhesions under myosin II tension, 
Curr. Biol. {\bf 20}, 1145 (2010).
 
\bibitem{ref:Wirtz_NatComm_2012}
S. I. Fraley, Y. Feng, A. Giri, G. D. Longmore, and D.Wirtz, 
Dimensional and temporal controls of three-dimensional cell migration by zyxin and binding partners, 
Nat. Commun. {\bf 3}, 719 (2012).

\bibitem{ref:Chabaud_NatComm_2015}
Chabaud, M. et al. 
Cell migration and antigen capture are antagonistic
processes coupled by myosin II in dendritic cells. 
Nature Commun. {\bf 6}, 7526 (2015).

\bibitem{ref:Levine_PRL_2013}
B.A. Camley, Y. Zhao, B. Li, H. Levine, and W.-J. Rappel, 
Periodic migration in a physical model of cells on micropatterns, 
Phys. Rev. Lett., {\bf 111}, 158102 (2013).

\bibitem{ref:Sens_PNAS_2020}
P. Sens, 
Stick-slip model for actin-driven cell protrusions, cell protrusions, cell polarization, and crawling, 
Proc. Natl. Acad. Sci. {\bf 117}, 24670 (2020). 

\bibitem{ref:Gov_PRR_2020}
J.E. Ron, P. Monzo, M.C. Gauthier, R. Voituriez, and N.S. Gov, 
One-dimensional cell motility patterns, 
Phys. Rev. Res., {\bf 2}, 033237 (2020).

\bibitem{ref:Prost_2007}
F. J\"{u}licher, K. Kruse, J. Prost, and J.-F. Joanny, 
Active behavior of the Cytoskeleton, 
Phys. Rep. \textbf{449},  3 (2007).

\bibitem{ref:Truskinovsky_PRL_2013}
P.~Recho, T.~Putelat, and L.~Truskinovsky,
Contraction-driven cell motility,
Phys. Rev. Lett. {\bf 111}, 108102 (2013).

\bibitem{ref:sekimoto1991}
K. Tawada, K. Sekimoto, 
Protein friction exerted by motor enzymes through a weak-binding interaction', 
 J. Theo. Biol.,  {\bf 150}, 193 (1991). 

\bibitem{ref:Recho_2016}
P.~Recho and L.~Truskinovsky, 
``Cell locomotion in one dimension,'' in {\em  Physical Models of Cell Motility}, p.~135, Springer, 2016.

\bibitem{putelat2018mechanical}
T.~Putelat, P.~Recho, and L.~Truskinovsky, 
Mechanical stress as a regulator of cell motility,'
Phys. Rev. E, {\bf 97}, 012410 (2018).

\bibitem{ref:Firtel_Cell_2003}
R. Meili and R.A. Firtel,
Two poles and a compass,
Cell, {\bf 114}, 153 (2003). 

\bibitem{maeda2008ordered}
Y.~T. Maeda, J.~Inose, M.~Y. Matsuo, S.~Iwaya, and M.~Sano, 
Ordered patterns of cell shape and orientational correlation during spontaneous cell migration,
PLoS one, {\bf 3}, e3734 (2008).

\bibitem{ref:Tsai_DevCell_2019}
T. Y.-C. Tsai, S.R. Collins, C.K. Chan, A. Hadjitheororou, P.-Y. Lam, S.S. Lou, H.W. Yang, J. Jorgensen, F. Ellett, D. Irimia, M.W. Davidson, R.S. Fischer, A. Huttenlocher, T. Meyer, J.E. Ferrell Jr, and J.A. Theriot,
Efficient front-rear coupling in neutrophil chemotaxis by dynamic myosin II localization,  
Dev. Cell {\bf 49}, 189 (2019).

\bibitem{nickaeen2017free}
M.~Nickaeen, I.~L. Novak, S.~Pulford, A.~Rumack, J.~Brandon, B.~M. Slepchenko,
  and A.~Mogilner,   
  A free-boundary model of a motile cell explains turning behavior,
  PLoS Comp. Biol, {\bf 13}, ~e1005862 (2017).

\bibitem{horton1995space}
G.~Horton and S.~Vandewalle, 
A space-time multigrid method for parabolic partial differential equations,
SIAM Journal on Scientific Computing, {\bf 16}, 848 (1995).

\bibitem{ref:SI}
See our Supplementary Material.

\bibitem{ref:Ingber_JCPhys_2006}
T.P. Lele, J. Pendse, S. Kumar, M. Salanga, J. Karavitis, and D.E. Ingber, 
Mechanical forces alter zyxin unbinding kinetics within focal adhesions of living cells, 
J. Cell Physiol., {\bf 207}, 187 (2006). 

\bibitem{ref:future}
J.-Y. Lo and H.-Y. Chen, manuscript in preparation. 

\bibitem{ref:Sano_BPJ}
H. Tanimoto and M. Sano, 
A simple force-motion relation for migrating cells revealed by multipole analysis of traction stress, 
Biophys. J., {\bf 106}, 16 (2014).

\bibitem{ref:Ohta_PhysicaD}
T. Ohta, M. Tarama, and M. SanoA simple model of cell crawling, 
Physica D, {\bf 318}, 3 (2016). 

\end{thebibliography}
\end{document}


\title{Supplementary material for \\ `` Mechanosensitive bonds induced complex cell motility patterns''}

\author{
  Jen-Yu Lo$^{1}$, Yuan-Heng Tseng$^{1}$ and Hsuan-Yi Chen $^{1,2,3}$
}
\affiliation{
  $^{1}$Department of Physics, National Central University, Jhongli 32001, Taiwan\\
  $^{2}$Institute of Physics, Academia Sinica, Taipei, 11529, Taiwan \\
  $^{3}$Physics Division, National Central for Theoretical Sciences, Taipei, 10617, Taiwan
}	 	

\date{\today}

\maketitle

\section{Dimensionless equations}
We introduce effective drag coefficient $\xi _{eff} = \xi + \alpha k_{on}/k_{0}$ and  choose $l_0 = \sqrt{\eta / \xi _{eff}}$ as the unit length, 
$t_0 =\eta / (\xi _{eff} D)$ as the unit time, $\sigma_0 = \xi _{eff} D$ as the unit stress, 
$n_0 = \xi /\alpha$ (notice that, here $\xi$, not $\xi _{eff}$ is used) as the unit density for adhesion complexes, 
and $c_0 = M/ \sqrt{\eta / \xi _{eff}}$ as the unit myosin concentration, where $M$ is the total number of myosin motors in the cell.
In the dimensionless form, the momentum equation is
\begin{equation}
    \frac{\partial^2 v}{\partial x^2} - \tilde{\xi} ( 1+n_b) v  = - \Tilde{\chi} \frac{\partial c}{\partial x},
    \label{eq:force_balance_dimensionless}
\end{equation}
myosin density evolution obeys
\begin{equation}
    \frac{\partial c}{\partial t} = \frac{\partial^2 c}{\partial x^2} - \frac{\partial (c v)}{\partial x},
    \label{eq:myosin_dimensionless}
\end{equation}
evolution of the density of the adhesion complexes is
\begin{equation}
    \frac{\partial n_b}{\partial t} 
    =- \Bar{k}_0 e^{- \Bar{k}_1 \partial _x v} n_b + \Bar{k}_{on}
    \label{eq:bonds_dimensionless}
\end{equation}
and the positions of the cell ends obey
\begin{equation}
        \frac{d l_{\pm}}{d t} = [v]_{l_{\pm}} \pm v_p^{\pm}.
\label{eq:lpm_dimensionless}
\end{equation}
The boundary condition for myosin current is 
\begin{equation}
        \left[c (v - \frac{dl}{dt}) - \frac{\partial c}{\partial x} \right]_{l_{\pm}} = 0,
        \label{eq:myosin_bc_dimensionless}
\end{equation}
and the stress continuity at the cell ends leads to 
\begin{equation}
    \sigma_{\pm} = - K \left(L - L_0 \right) = \left[\frac{\partial v}{\partial x}\right]_{l_{\pm}} + \Tilde{\chi} c_{l_{\pm}}.
    \label{eq:stress_bc_dimensionless}
\end{equation}
Here all variables and $x$, $t$ are dimensionless.
The definitions of the parameters and important physical quantities are listed in Table~S1.  The dimensionless parameters in our model are listed in Table~S2. 

\begin{table}[htb]
\renewcommand\arraystretch{1.5}
\centering
\begin{tabular}[t]{llr}
\hline
\hline
&Physical meaning                             &  Symbol\\
\hline

&effective drag coefficient                 & $\xi _{eff} = 
                                                                                      \xi + \alpha k_{on}/k_0$ \\ 
&unit length                                       &  $l_0=\sqrt{\eta / \xi _{eff}}$\\
&unit time                                          &  $t_0=\eta / \xi _{eff} D$\\
&unit stress                                        &  $\sigma_0 = \xi _{eff} D$\\
&unit myosin motors concentration   &  $c_0=\frac{M}{\sqrt{\eta / \xi _{eff}}}$\\
&unit density of cell-substrate bonds &  $n_0 = \xi /\alpha$\\
\hline
\hline
\end{tabular}
\caption{Definitions of physical parameters and characteristic quantities in our model.}
\label{Table1}
\end{table}

\begin{table}[htb]
\renewcommand\arraystretch{1.5}
\centering
\begin{tabular}[t]{llr}
\hline
\hline
&Physical meaning                             &  Symbol\\
\hline
& unbinding rate                                 &  $\bar{k}_0 = t_0 k_0$ \\
& coefficient for strain-rate-dependent unbinding        &  $\bar{k}_1 = k_1/t_0$ \\
& binding rate                                     &  $\bar{k}_{\rm on} = k_{\rm on}t_0 / n_0$ \\
& base actin polymerization speed &  $\bar{v}_p^{(0)} = v_p^{(0)} l_0 /t_0$ \\
& coefficient for stress-dependent actin polymerization  & $\bar{v}_p^{(1)} = v_p^{(1)} l_0$\\
& coefficient for cell polarization effect on actin polymerization &$\bar{v}_p^{(2)} = v_p^{(2)} t_0/l_0$\\
& contractility                               &  $\Tilde{\chi}=c_0 \chi / \sigma_0$\\
& cell elastic constant                   &  $K=\gamma l_0/ \sigma_0$\\
& drag coefficient                         &  $\tilde{\xi} = \xi /(\xi + \alpha k_{\rm on}/k_0)$ \\
\hline
\hline
\end{tabular}

\caption{Definitions of dimensionless parameters in our model.}
\label{Table2}
\end{table}%

\section{Numerical method and choice of parameters}
In our numerical scheme, each iteration updates all dynamical variables by integrating the evolution equations over a small time interval $\Delta t$ with a finite difference method.   
First, $[v]_{l_{\pm}}$ and $v_p^{\pm}$ from the previous iteration were substituted into Eq.~(\ref{eq:lpm_dimensionless}) to obtain the new positions of the cell ends. 
The densities of the adhesion complexes and myosin motors are updated from the flow field of the previous iteration by integrating the evolution Eq.~\eqref{eq:bonds_dimensionless} of $n_b$ and the myosin advection-diffusion Eq.~\eqref{eq:myosin_dimensionless}.
The force balance  Eq.~\eqref{eq:force_balance_dimensionless} with the updated bond density and myosin concentration is then solved to obtain the new flow field.

The numerics were carried out by dividing the cell into $N_x = 100$ segments and approximating the spatial derivatives by the finite difference method with the size of a time step $\Delta t={10}^{-6}$. 
The typical material parameters are $D$ $\sim$  0.025 $\mu {\rm m}^2 / s$ \cite{ref:Luo_BPJ_2012}, 
 $l_0 \sim 10$ $\mu {\rm m}$ and unit time $t_0 \sim 10^3$~s~\cite{ref:Barnhart_PLoSBiol_2011}. $k_0$ is the rate of dissociation of mature focal adhesion and its typical value $\sim$ $1/{\rm min}$ \cite{ref:Wang_JCB_1984}, and we chose $\bar{k}_0=3$. $\bar{k}_{\rm on}$ is chosen to be 6 for most simulations and the dimensionless density of bonds in the absence of mechanosensitivity is of order unity.  
The dimensionless natural length of the cell $L_0 = 1$ for a typical cell. 
This means that in the absence of adhesion complexes, drag and viscous forces are of the same order of magnitude. 
The dimensionless total number of myosin motors in the cell $c_0 L_0 = 1$.   
\vskip 3 in

\newpage
\section{Simulation results}
The motility phase diagrams in the plane spanned by $\tilde{\chi}$ and $k_1$ are presented in the main text.
Here Fig.~S1 shows the phase diagram in the plane spanned by $v_p^{(0)}$ and $k_1$.  From this figure we can see how the actin polymerization rate affects the motility behavior of the cell.  
It is clear that the moving state of a cell with small $k_1$ is that with a constant velocity, while the moving state of a cell with large $k_1$ is periodic back-and-forth.  This is in agreement with the phase diagrams in the main text.  
Furthermore, there are several complex motility patterns between these two states, including stick-slip motion 
and behaviors that can be seen as combinations of back-and-forth and stick-slip movement. 
The trajectories for stick-slip movement and periodic back-and-forth movement with stick-slip are 
shown in Fig.~1(c)(d) of the main text.  
The trajectories for zig-zag movement with stick-slip  
and double-period back-and-forth motion are shown in Fig.~4(a)(b) of the main text. 
\begin{figure}[!h]
 \centering 
\includegraphics[width=0.75 \linewidth]{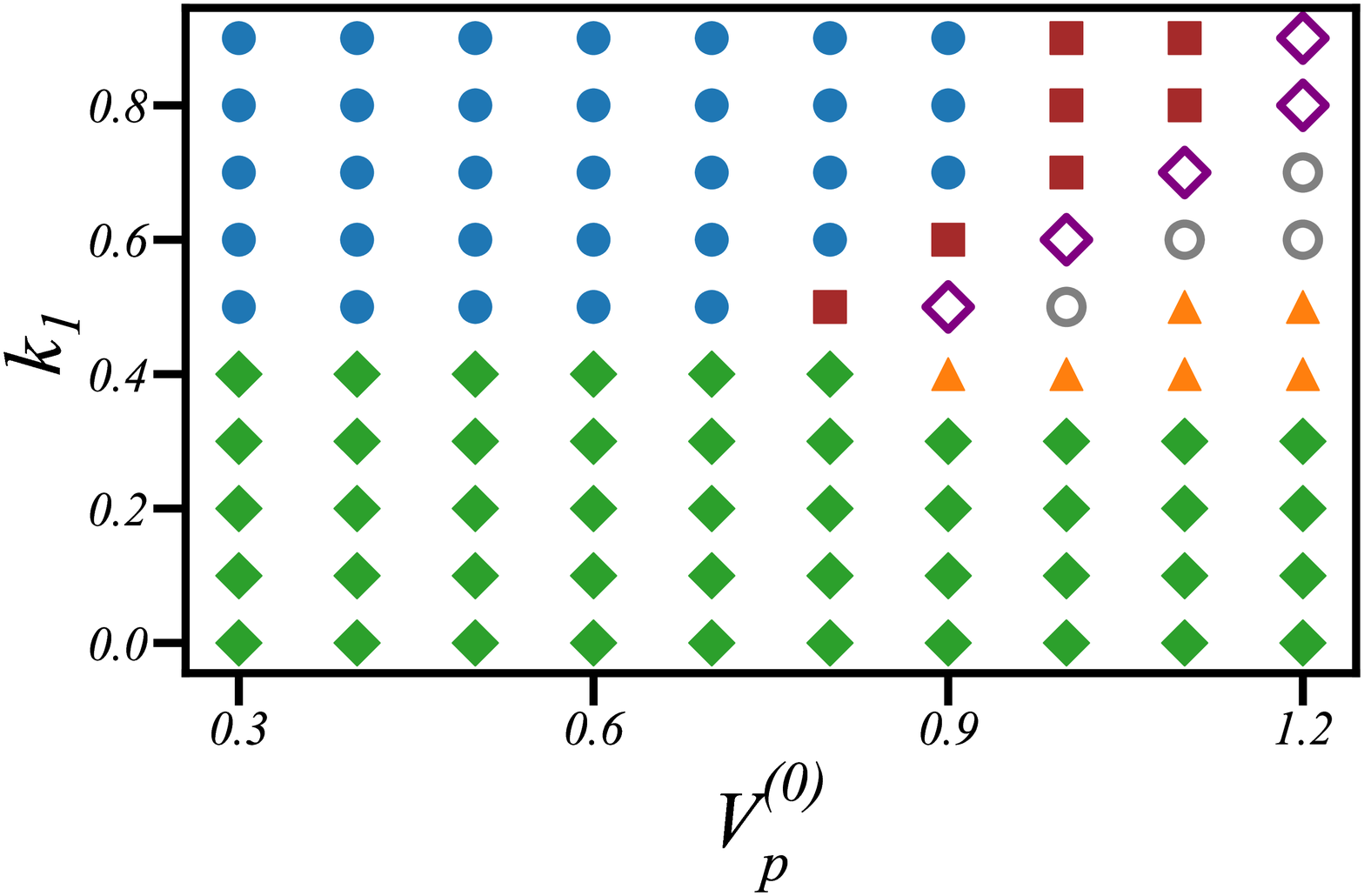}
\label{fig:phase_diagrams}
 \caption[S1]{Phase diagram
 for the motility behavior predicted by the reduced model.  $K = 100$, $\tilde{\xi} = 1/3$, $k_{\rm on}= 6$, $k_0=3$, 
$\tilde{\chi}=14$, $v_p^{(1)}=0.5$, $v_p^{(2)}=0.2$.
The cells show the following motility behaviors: 
constant velocity motion (green diamonds), 
periodic back-and-forth movement (blue circles), 
stick-slip movement (orange triangles), 
periodic back-and-forth movement with stick-slip (empty gray circles), 
zig-zag movement with stick-slip (empty purple diamonds), 
and double-period back-and-forth motion (brown squares). 
 }
\end{figure}
\vskip 1 in

Figure S2 shows the distribution of adhesion complexes and myosin motors for a cell that undergoes stick-slip movement and periodic back-and-forth movement. 
Similar to the distributions shown in Fig.~2 of the main text, 
for a moving cell, myosin motors are always located relatively close to the trailing end, and the density of adhesion complexes is always higher in a region close to the leading end.  
\newpage
\begin{figure}[!h]
\mbox{
\subfigure[]{\includegraphics[width=0.9\linewidth]{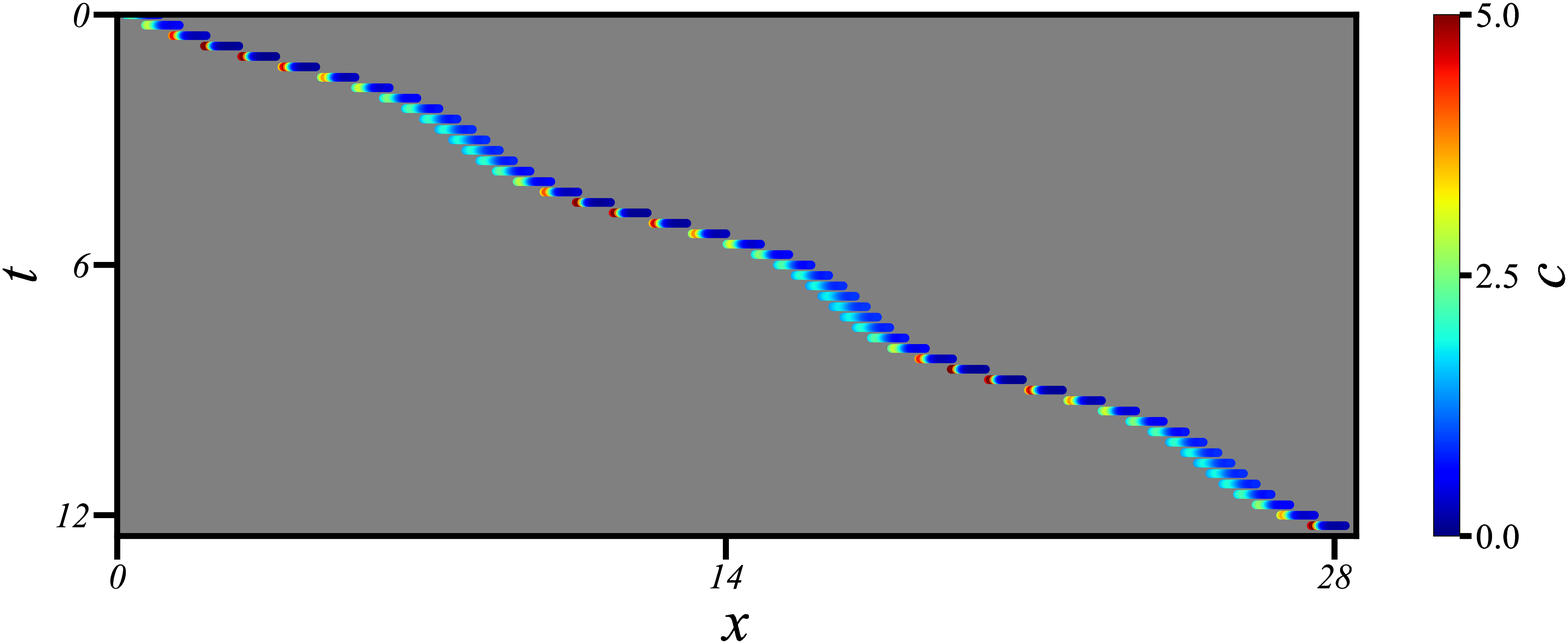}}
}
\mbox{
\subfigure[]{\includegraphics[width=0.9\linewidth]{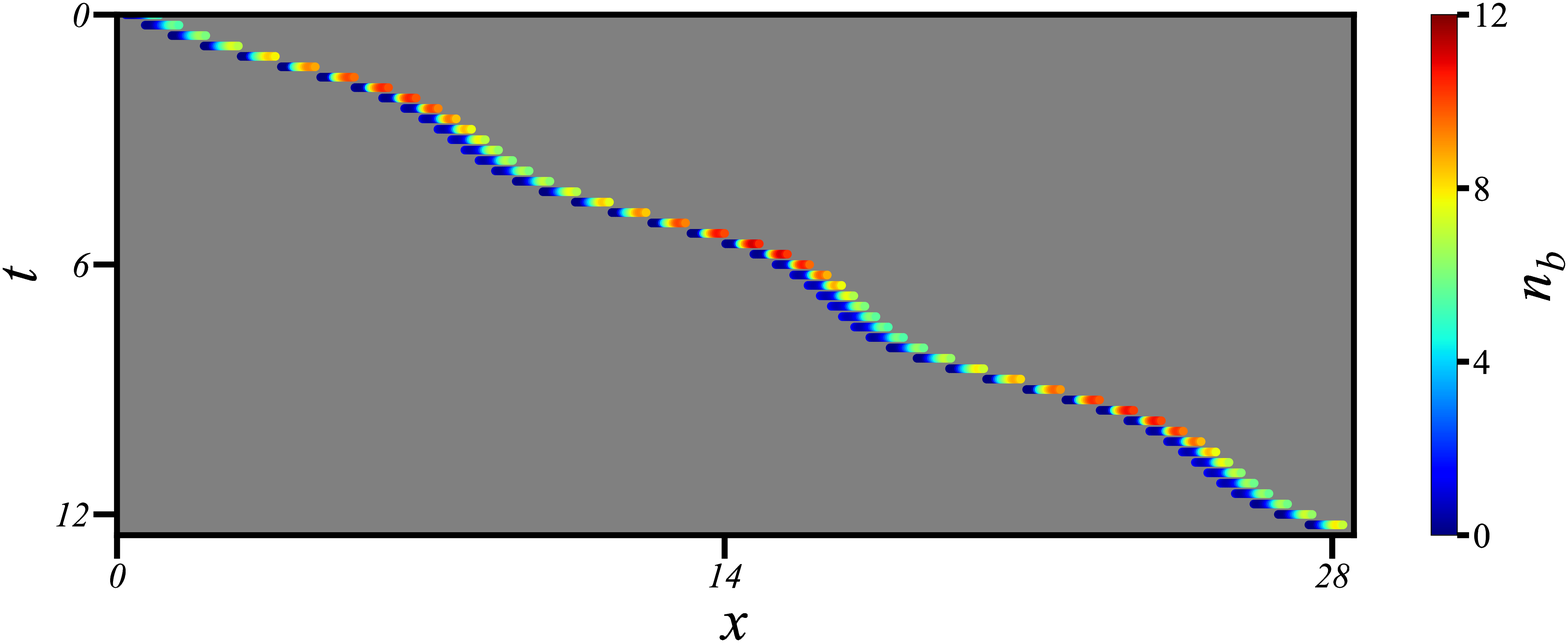}}
}
\mbox{
\subfigure[]{\includegraphics[width=.4\linewidth]{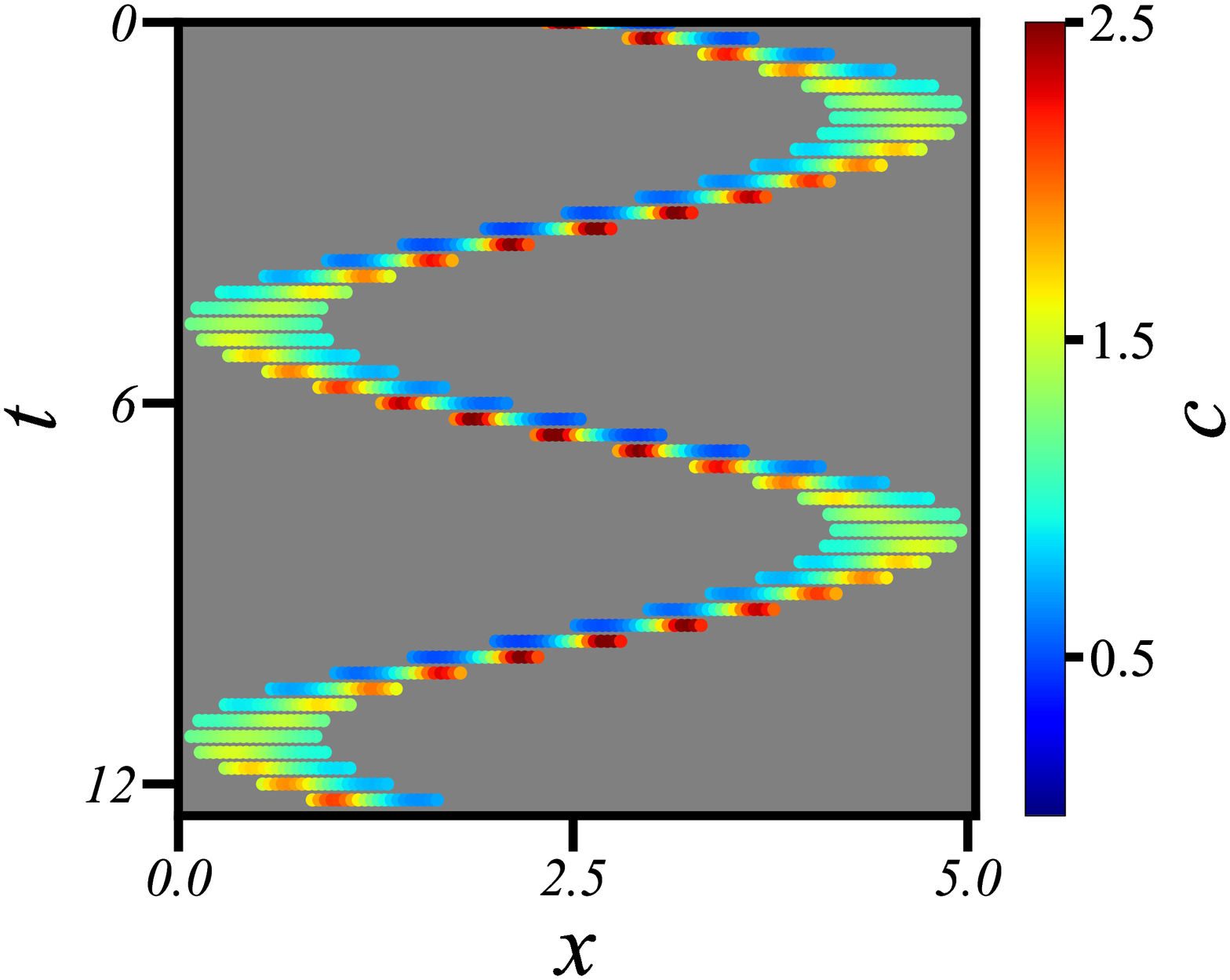}}
}
\mbox{
\subfigure[]{\includegraphics[width=.4\linewidth]{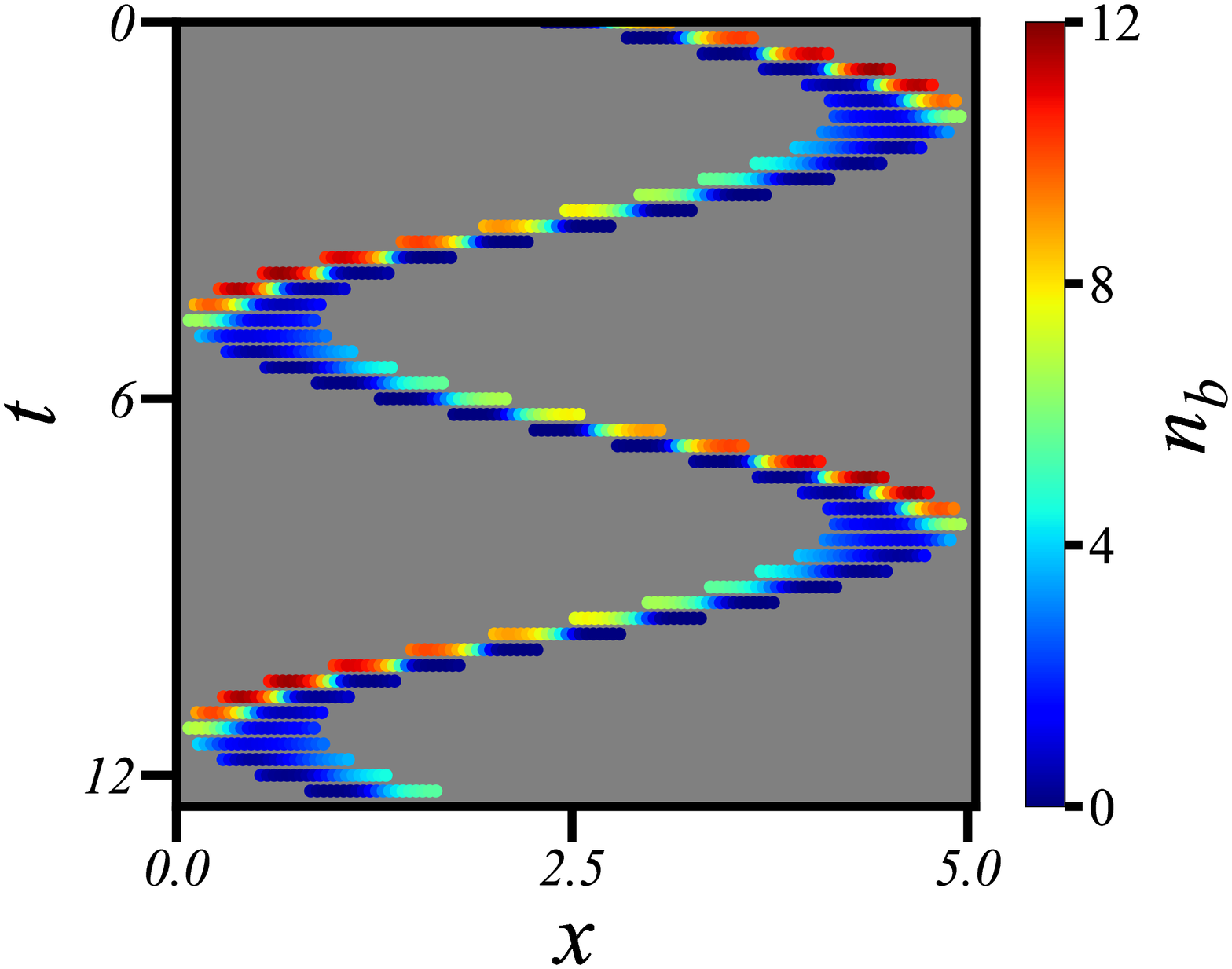}}
}
\label{fig:distributions}
 \caption{Distribution of adhesion complexes and myosin motors for  $K = 100$, $\tilde{\xi} = 1/3$, $k_{\rm on}= 6$, $k_0=3$, $v_p^{(0)}=0.2$,  $v_p^{(1)}=0.5$, 
               $v_p^{(2)}=2$, and  (a)(b) $k_1=0.12$, $\tilde{\chi} = 18$, (c)(d) $k_1 = 0.25$, $\tilde{\chi} = 17$.
 }
\end{figure}

\newpage
\section{The simplified model}
\label{App:simplified_model}
Since the adhesion complexes are concentrated close to the cell ends, to simplify the analysis, we assume
\begin{eqnarray}
n_b(x,t) = N_{f} \delta (x-x_f) + N_{b} \delta (x-x_b),
\end{eqnarray}
where $x_{f} = l_+ - \epsilon$, $x_{b} = l_- + \epsilon$, and $\epsilon$ is a very small length. 

\subsection{Stress field and flow field}
Because there are no adhesion complexes in $x_{b} < x < x_{f}$, the dimensionless momentum equation in this region is  
\begin{eqnarray}
\partial _x \sigma = \tilde{\xi} v, \ \  \sigma = \partial _x v + \tilde{\chi} c, \ \ x_{b} < x < x_{f}.
\label{eq:stress_simplified}
\end{eqnarray}
Integrating the full momentum equation from $l_{+(-)}$ to $x_{f (b)}$, we find
\begin{eqnarray}
\sigma _{f} &=& -K(l-l_0) - \tilde{\xi} N_{f} v_{f},  \nonumber \\
\sigma _{b} &=& -K(l-l_0) + \tilde{\xi} N_{b} v_{b}.  
\end{eqnarray}
Here $v_f = v(x_f,t) \approx dl_+/dt - v_p^+$, and $v_b = v(x_b,t) \approx dl_-/dt + v_p^-$.
The solution of stress from these equations is~\cite{ref:Recho_2016}
\begin{eqnarray}
\sigma (x,t) = \sigma _{f} \frac{\sinh \left[\tilde{\xi}^{1/2}( x-x_b) \right]}{\sinh \left[\tilde{\xi}^{1/2}( x_f-x_b) \right]} +  \sigma _{b} \frac{\sinh \left[\tilde{\xi}^{1/2}( x_f-x) \right]}{\sinh \left[\tilde{\xi}^{1/2}( x_f-x_b) \right]}
                     + \tilde{\chi} \tilde{\xi}^{1/2} \int _{x_b}^{x_f} G(x,x')c(x',t)dx', \nonumber \\
\end{eqnarray}
where 
\begin{eqnarray}
G (x,x') = \frac{\sinh \left[\tilde{\xi}^{1/2}( x_f-x) \right]\sinh \left[\tilde{\xi}^{1/2}( x'-x_b) \right]}{\sinh \left[\tilde{\xi}^{1/2}( x_f-x_b) \right]} - \Theta (x'-x)\sinh \left[\tilde{\xi}^{1/2}( x'-x) \right],
\end{eqnarray}
$\Theta (x) $ is the Heaviside step function. 
This leads to the following expression for the flow field, 
\begin{eqnarray}
v(x,t) = \frac{1}{\tilde{\xi}^{1/2}}\left\{
                                                     \sigma _{f} \frac{\cosh \left[\tilde{\xi}^{1/2}( x-x_b) \right]}{\sinh \left[\tilde{\xi}^{1/2}( x_f-x_b) \right]} -  \sigma _{b} \frac{\cosh \left[\tilde{\xi}^{1/2}( x_f-x) \right]}{\sinh \left[\tilde{\xi}^{1/2}( x_f-x_b) \right]}
                                                    + \tilde{\chi}  \int _{x_b}^{x_f} \partial _xG(x,x')c(x',t)dx'
                                                    \right\}.
\nonumber \\
\label{eq:velocity_simplified}
\end{eqnarray}

\subsection{Myosin concentration}
Substituting Eq.~(\ref{eq:velocity_simplified}) into the advection-diffusion for myosin concentration, one obtains the following equation which does not have an explicit dependence on the velocity field.
\begin{eqnarray}
 \partial _t c(x,t) 
= D \partial _x^2 c 
                               -  \frac{1}{\tilde{\xi}^{1/2}}  &\partial _x& \left\{
                                                     \left[ 
                                                     \sigma _{f} \frac{\cosh \left[\tilde{\xi}^{1/2}( x-x_b) \right]}{\sinh \left[\tilde{\xi}^{1/2}( x_f-x_b) \right]} -  \sigma _{b} \frac{\cosh \left[\tilde{\xi}^{1/2}( x_f-x) \right]}{\sinh \left[\tilde{\xi}^{1/2}( x_f-x_b) \right]} \right] c(x,t)  \right. \nonumber \\
                              &&    \left.     + \tilde{\chi}  \int _{x_b}^{x_f} \  c(x,t) \partial _xG(x,x')c(x',t)dx'
                                                    \right\}.
\nonumber \\
\label{eq:c_simplified}
\end{eqnarray}

\subsection{Velocity and length of the cell}
    The velocity of the cell $V_{\rm cell} = \frac{1}{2}\left( \frac{dl_+}{dt} + \frac{dl_-}{dt}\right) $ is 
\begin{eqnarray}
V_{\rm cell} 
                    &=& \frac{1}{2 \tilde{\xi}^{1/2}}  \frac{\cosh \left(\tilde{\xi}^{1/2} L\right) + 1}{\sinh \left(\tilde{\xi}^{1/2} L\right)} (\sigma _f - \sigma _b)
                             +  \frac{\tilde{\chi}}{2}  \int _{x_b}^{x_f} \frac{\sinh  \left[\tilde{\xi}^{1/2} (x_f-x')\right] - \sinh  \left[\tilde{\xi}^{1/2} (x'-x_b)\right]}{\sinh \left(\tilde{\xi}^{1/2} L\right)} c(x',t)dx'  \nonumber \\
                    &&   + \frac{v_p^+ - v_p^-}{2}.
\label{eq:V_cell_simplified}
\end{eqnarray}
The evolution of the length of the cell $\frac{dL}{dt}= \frac{dl_+}{dt} - \frac{dl_-}{dt}$ obeys
\begin{eqnarray}
\frac{dL}{dt}
                    &=& \frac{1}{\tilde{\xi}^{1/2}} \frac{\cosh \left(\tilde{\xi}^{1/2} L\right)-1}{\sinh \left(\tilde{\xi}^{1/2} L\right)} (\sigma _f + \sigma _b)
                             -  \tilde{\chi}  \int _{x_b}^{x_f} \frac{\sinh  \left[\tilde{\xi}^{1/2} (x_f-x')\right] + \sinh  \left[\tilde{\xi}^{1/2} (x'-x_b)\right]}{ \sinh \left(\tilde{\xi}^{1/2} L\right)} c(x',t)dx' \nonumber \\
                     && + (v_p^+ + v_p^-). \nonumber \\
\end{eqnarray}

\subsection{Symmetries of the system}
It is helpful to introduce the following variables
\begin{eqnarray}
N &=& N_f + N_b, \ \ \Delta N = N_f - N_b, \nonumber \\
v_p &=& \frac{v_p^+ + v_p^-}{2}, \ \ \Delta v_p = v_p^+ - v_p^-, \nonumber \\
\sigma _S &=& \frac{\sigma _f + \sigma _b}{2} = - K(L-L_0) - \frac{\tilde{\xi}}{2} \left[ N \left( \frac{dL/dt}{2}-v_p\right) + \Delta N \left( V_{\rm cell} - \frac{\Delta v_p}{2} \right)\right] , \nonumber \\
\sigma _A &=& \frac{\sigma _f - \sigma _b}{2} = - \frac{\tilde{\xi}}{2} \left[ N \left(  V_{\rm cell} - \frac{\Delta v_p}{2} \right) + \Delta N \left( \frac{dL/dt}{2}-v_p\right) \right], 
\label{eq:symmetry}
\end{eqnarray}
and 
\begin{eqnarray}
y &=& x - \frac{l_+ + l_-}{2}.  
\end{eqnarray}
Notice that 
 $dL/dt$, $N$, $v_p$, and $\sigma _S$ are symmetric under spatial inversion ($y \rightarrow -y$), 
 while $V_{\rm cell}$, $\Delta N$, $\Delta v_p$, and $\sigma _A$ are antisymmetric under spatial inversion. 

The velocity of the cell $V_{\rm cell}$ can be expressed in terms of these parameters and variables that have clear parity signatures.  
First, notice that  
only the part of $c(y)$ that is anti-symmetric under $y  \rightarrow -y$ contribute to the $\tilde{\chi}$-dependent term of Eq.~(\ref{eq:V_cell_simplified}), 
and for a slow-crawling cell this part should be significant only in the small $|y|$ region.  
Thus by expanding the $\tilde{\chi}$-dependent term of $V_{\rm cell}$ to the leading order in  $y$, one finds that 
\begin{eqnarray}
V_{\rm cell}=
 \frac{1}{2} \left[ 
                            \Delta v_p + \frac{2}{\tilde{\xi}^{1/2}}\frac{\cosh \left(\tilde{\xi}^{1/2}L\right) + 1}{\sinh \left(\tilde{\xi}^{1/2}L \right)} \sigma _A 
                             - \tilde{\chi}\tilde{\xi}^{1/2}\ \frac{\cosh \left(2 \tilde{\xi}^{1/2}L\right)}{\sinh \left(\tilde{\xi}^{1/2}L\right)} y_c + ...
                   \right], \nonumber 
\end{eqnarray}
where
\begin{eqnarray}
y_c \equiv \int _{-L/2}^{L/2}  y\ c(y,t) \ dy,
\end{eqnarray}
and  ``...'' represents terms of higher order in this expansion.  
The expression for $V_{\rm cell}$ can be further simplified by taking  $\Delta v_p \approx \beta V_{\rm cell}$ ($\beta$ is independent of $V_{\rm cell}$) for a slow crawling cell and substituting Eq.~(\ref{eq:symmetry}) for $\sigma _A$.  
Finally, one obtains
\begin{eqnarray}
V_{\rm cell} &=& - \frac{\tilde{\xi}^{1/2}}{1 - \beta/2} \left[ 1 + \frac{ \tilde{\xi}^{1/2}}{2}\frac{\cosh \left(  \tilde{\xi}^{1/2}L\right) +1}{\sinh \left( \tilde{\xi}^{1/2}L \right) }\ N\right]^{-1} \nonumber \\
                     & & \times   \left[  \frac{1}{2} \frac{\cosh \left(  \tilde{\xi}^{1/2}L\right) +1}{\sinh \left(  \tilde{\xi}^{1/2}L\right) } \left( \frac{1}{2}\frac{dL}{dt} - v_p \right) \Delta N
                              +  \tilde{\chi} \frac{\cosh \left(  \tilde{\xi}^{1/2}L/2\right) }{\sinh \left(  \tilde{\xi}^{1/2}L\right) } y_c + ... 
                       \right]. 
\end{eqnarray}
This expression tells us that $V_{cell}$ is nonzero only when $\Delta N$ (asymmetry in the distribution of adhesion complexes) or $y_c$ (asymmetry in the distribution of myosin motors) is nonzero.  

Similar calculation leads to the following expression for the evolution of the length of the cell,
\begin{eqnarray}
\frac{dL}{dt} = 2 v_p &+& \left[1+\frac{\tilde{\xi}^{1/2}}{2}\frac{\cosh (\tilde{\xi}^{1/2}L ) -1}{\sinh (\tilde{\xi}^{1/2}L)} \ N\right]^{-1}  \nonumber \\
&\times &
\left\{ \tilde{\xi}^{-1/2}\frac{\cosh (\tilde{\xi}^{1/2}L ) -1}{\sinh (\tilde{\xi}^{1/2}L)} \left[
-2 K (L-L_0) +  \Delta N \left(1-\frac{\beta}{2}\right) V_{\rm cell} \right] 
                       - 2 \tilde{\chi}\ \frac{\sinh \left(  \tilde{\xi}^{1/2}L/2\right) }{\sinh \left( \tilde{\xi}^{1/2}L\right) } C_{\rm tot}\right\}, \nonumber \\
\end{eqnarray}
where 
\begin{eqnarray}
C_{\rm tot}  \equiv \int _{-L/2}^{L/2} \ c(y,t) dy \equiv 1
\end{eqnarray}
is the total amount of myosin motors in the cell, which is unity in our dimensionless expression.  
Note that all terms on the right-hand side of $dL/dt$ are even under $y \rightarrow - y$.  

\subsection{The simplified model and bifurcations}
From the previous analysis, it is clear that 
symmetry under $y \rightarrow -y$ plays an important role in $V_{\rm cell}$ and $dL/dt$.  
Based on these observations, a simplified model is proposed for slow-crawling cells. 

First, the evolution equations of $N_f$ and  $N_b$ are
\begin{eqnarray}
\frac{d N_f}{dt} = k_{\rm on} - (k_{\rm off}^{(0)} + k_{\rm off}^{(1)} \ y_c ) N_f , \nonumber \\
\frac{d N_b}{dt} = k_{\rm on} - (k_{\rm off}^{(0)} - k_{\rm off}^{(1)} \ y_c ) N_b . \nonumber
\end{eqnarray}
This leads to 
\begin{eqnarray}
\frac{dN}{dt} &=&  2 k_{\rm on} - k_{\rm off} ^{(0)} \ N - k_{\rm off}^{(1)} \ y_c \ \Delta N  , \nonumber \\ 
\frac{d\Delta N}{dt} &=& - k_{\rm off}^{(0)} \Delta N - k_{\rm off}^{(1)} \ N y_c.
\label{eq:dNdtdDeltaNdt}
\end{eqnarray}
We expect $k_{\rm off}^{(1)} >0$ because a cell moving at constant velocity in the $+x$ direction should have $y_c <0$ and $\Delta N >0$.  

Next, the following evolution equation for $y_c$ is proposed
 \begin{eqnarray}
\frac{dy_c}{dt} = - \Gamma \left[ - (\tilde{\chi} - \tilde{\chi}_c ) y_c -  a_{\Delta N} \Delta N + a_3 y_c^3 \right].
\label{eq:dycdt}
\end{eqnarray}
The above equation describes a cell that becomes polarized ($y_c \neq 0$) when $\tilde{\chi}$ is sufficiently large.  Furthermore, $a_{\Delta N}$ tells us how nonzero $\Delta N$ affects the evolution of $y_c$.  
In general, $\tilde{\chi}_c$, $a_{\Delta N} $,  and $a_3$ all depend on $L$ and $N$.  $a_3 >0$ such that $y_c$ remains finite.

To focus on the physics that are most relevant to the transitions between different motility behaviors, we 
neglect the $N$-dependencies in $\tilde{\chi}_c$ and $a_{\Delta N}$ 
as they do not change the symmetry properties of the evolution equation of $y_c$.  
This approximation is expected to be suitable for slow-moving cells.  
Furthermore, the $L$-dependencies of all coefficients in our simplified model can be neglected by considering the large-$K$ regime such that 
$L \rightarrow L_0$.
In this regime, 
\begin{eqnarray}
V_{\rm cell} &=& \frac{\tilde{\xi}^{1/2}}{(1-\frac{\beta}{2})\left[ 1 + \frac{\tilde{\xi}^{1/2}}{2}\frac{\cosh (\tilde{\xi}^{1/2}L_0) + 1}{\sinh (\tilde{\xi}^{1/2} L_0)}N\right] }
                          \left\{ \left[ \frac{1}{2}\frac{\cosh (\tilde{\xi}^{1/2}L_0) + 1}{\sinh (\tilde{\xi }^{1/2} L_0)} v_p \right] \Delta N - \left[\tilde{\chi}\frac{\cosh (\tilde{\xi}^{1/2}L_0/2)}{\sinh (\tilde{\xi}^{1/2}L_0)}\right]y_c  \right\} \nonumber \\
                       &\equiv& \frac{1}{1-\beta/2}(\lambda _{\nu_1 v_p}\Delta N - \tilde{\chi} \lambda_{\nu_2}y_c).
\end{eqnarray}
Here $\lambda _{\nu_1 v_p}$ and $ \lambda_{\nu_2}$ are $N$-dependent parameters.

Many interesting features of the system described by Eqs.~(\ref{eq:dNdtdDeltaNdt})(\ref{eq:dycdt}) can be studied analytically.
First, the steady-state solutions include the rest-state solution
\begin{eqnarray}
\Delta N = y_c = 0, \ N = \frac{2k_{\rm on}}{k_{\rm off}^{(0)}} \equiv N_0, \ \ \ 
\end{eqnarray}
 and solutions for a cell moving in the $\pm$ x-direction with a constant velocity
\begin{eqnarray}
     y_c &=& \mp \sqrt{\frac{p_4 + p_1^2 p_2 - \sqrt{(p_4 + p_1^2 p_2)^2 - 4 p_1^2p_4 (p_2-p_1p_3N_0)}}{2p_1^2p_4}},  \nonumber  \\
    \Delta N &=& - \frac{p_1N_0}{1 +\left( p_1 y_c\right)^2}y_c,   \nonumber \\
    N &=& \frac{N_0}{1- \left( p_1y_c\right)^2},
\end{eqnarray}
where $p_1 = k_{\rm off}^{(1)}/k_{\rm off}^{(0)}$, $p_2 = \Gamma (\tilde{\chi} - \tilde{\chi}_c)/{k_{\rm off}^{(0)}}$, $p_3 = \Gamma a_{\Delta N}/{k_{\rm off}^{(0)}}$, and $p_4 = \Gamma a_3/{k_{\rm off}^{(0)}}$. 
Further checking the linear stability of the rest state shows that the transition from the rest state to the state with constant velocity is a pitchfork bifurcation.  On the other hand,
the transition from the rest state to the periodic back-and-forth movement is a Hopf bifurcation: 
Pitchfork bifurcation (rest/constant-velocity transition)  happens when 
\begin{eqnarray}
\tilde{\chi} = \tilde{\chi}_c + 2 a_{\Delta N}\frac{k_{\rm on} k_{\rm off}^{(1)}}{(k_{\rm off}^{(0)})^2}
\end{eqnarray} 
and 
\begin{eqnarray}
\Gamma (\tilde{\chi} - \tilde{\chi}_c)-k_{\rm off}^{(0)} <0.
\end{eqnarray} 
Hopf bifurcation (rest/back-and-forth-motion transition) occurs when  
\begin{eqnarray}
\tilde{\chi} = \tilde{\chi}_c +  k_{\rm off}^{(0)}/\Gamma
\end{eqnarray}
and 
\begin{eqnarray}
\tilde{\chi}-\left[\tilde{\chi}_c + 2 a_{\Delta N}\frac{k_{\rm on} k_{\rm off}^{(1)}}{(k_{\rm off}^{(0)})^2}\right] <0.
\end{eqnarray} 
The phase diagram for the motility behavior predicted by this phenomenological model is shown in Fig.~S3.  It is qualitatively similar to the phase diagrams of the active gel model. 
The differences are likely due to the approximations we made when constructing the simplified model. 
 For example, assuming a constant cell length and assuming that the dynamics of $y_c$ are independent of $N$ and $L$ are likely to have some effects on the detailed shape of the phase boundaries. 
\begin{figure}[!h]
 \centering 
\includegraphics[width=0.6\linewidth]{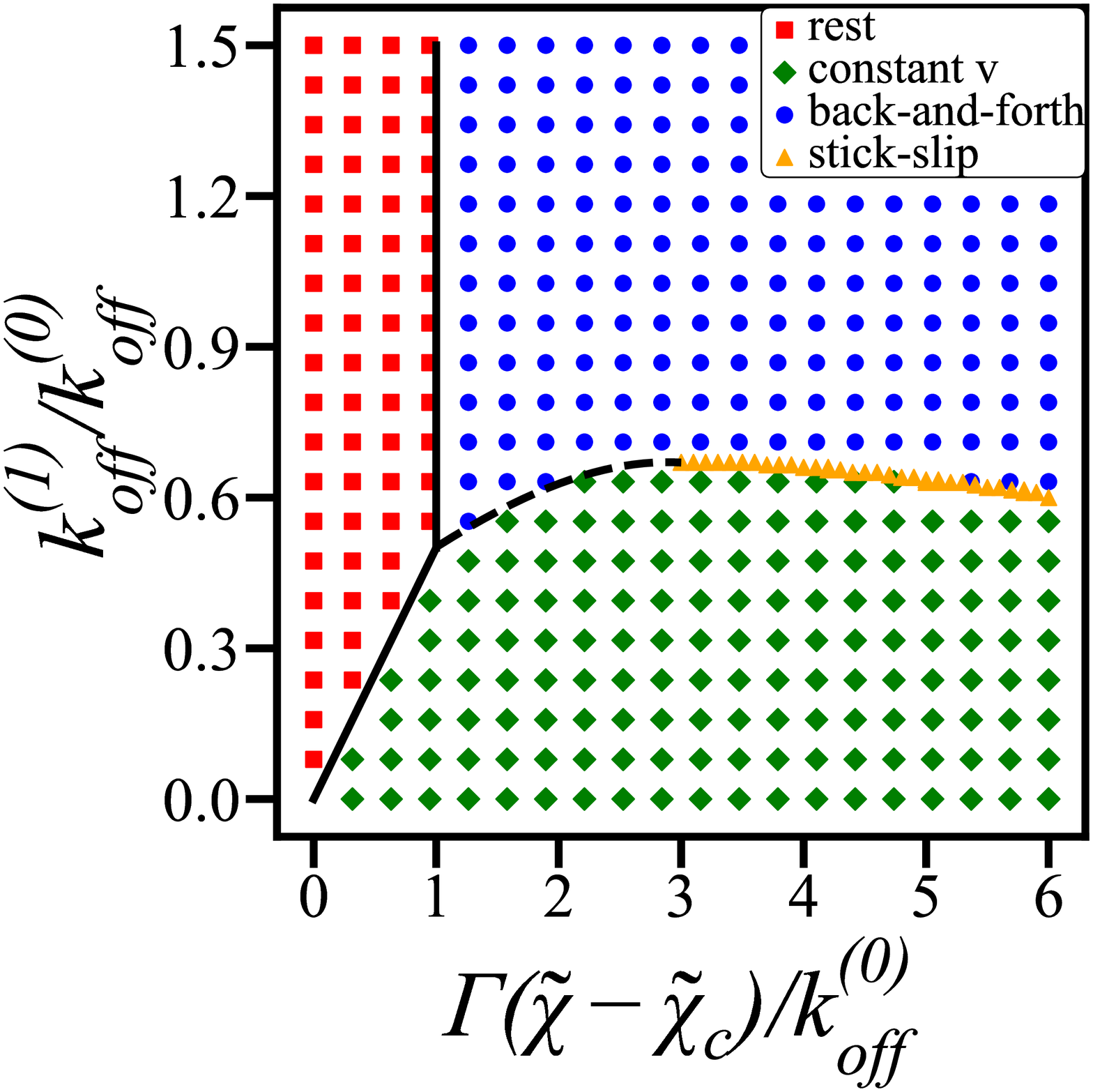}\label{fig:phase_diagram_simplified_model}
 \caption{Phase diagram for the motility behavior predicted by the simplified model.  
 The following motility patterns are found: 
 a cell at rest (red squares), 
 a cell moving at constant velocity (green diamonds), 
 a cell performs stick-slip movement (orange triangles),
 a cell performs back-and-forth movement with stick-slip (at $k^{(1)}_{\rm off}/k^{(0)}_{\rm off}$ slightly greater than
 those orange triangles so that we cannot show), 
 and a cell performs periodic back-and-forth movement (blue circles). 
 The boundary between the rest and constant velocity movement is  $\tilde{\chi} = \tilde{\chi}_c + 2 a_{\Delta N}\frac{k_{\rm on} k_{\rm off}^{(1)}}{\left(k_{\rm off}^{(0)}\right)^2}$.  
 The boundary between the rest and periodic back-and-forth movement states  is 
 $\tilde{\chi} = \tilde{\chi}_c +  k_{\rm off}^{(0)}/\Gamma$.    }
\end{figure}